\documentclass[preprint,prd,tightenlines,floatfix,
nofootinbib,eqsecnum,superscriptaddress]{revtex4-1}

\usepackage[T1]{fontenc}		

\usepackage{amsmath,amsfonts,amssymb,amstext,mathrsfs}
\usepackage{mathpazo}

\usepackage[dvips]{graphicx}
\usepackage{epsf,float}
\usepackage{revsymb}

\usepackage{dcolumn}
\usepackage{braket}
\usepackage{color,xcolor}
\usepackage{graphicx}
\usepackage{subfigure}
\usepackage{multirow}
\usepackage{tabularx}
\usepackage{pstricks}
\usepackage[section]{placeins}
\usepackage{booktabs}
\usepackage{array}

\usepackage{tablefootnote}

\usepackage{commath}

\usepackage{hyperref}

\usepackage{lineno}

\newcommand{\Pom}{\mathbb{P}}

\newcommand{\bk}{\mbox{\boldmath $k$}}
\newcommand{\bpa}{\mbox{\boldmath $p_{a}$}}

\newcommand{\bhpi}{\mbox{\boldmath $p_{i \perp}'$}}
\newcommand{\bhpii}{\mbox{\boldmath $p_{1 \perp}'$}}
\newcommand{\bhpjj}{\mbox{\boldmath $p_{2 \perp}'$}}
\newcommand{\khpi}{\mbox{\boldmath $k_{\perp}$}}

\usepackage[normalem]{ulem}  

\begin{document}

\title{\boldmath 
Exclusive diffractive bremsstrahlung of one and two photons \\
at forward rapidities:
possibilities for experimental studies \\
in $pp$ collisions at the LHC}

\vspace{0.6cm}

\author{Piotr Lebiedowicz}
\email{Piotr.Lebiedowicz@ifj.edu.pl}
\affiliation{Institute of Nuclear Physics Polish Academy of Sciences, Radzikowskiego 152, PL-31342 Krak{\'o}w, Poland}

\author{Otto Nachtmann}
\email{O.Nachtmann@thphys.uni-heidelberg.de}
\affiliation{Institut f\"ur Theoretische Physik, Universit\"at Heidelberg,
Philosophenweg 16, D-69120 Heidelberg, Germany}

\author{Antoni Szczurek}
\email{Antoni.Szczurek@ifj.edu.pl}
\affiliation{Institute of Nuclear Physics Polish Academy of Sciences, Radzikowskiego 152, PL-31342 Krak{\'o}w, Poland}
\affiliation{College of Natural Sciences, 
Institute of Physics, University of Rzesz{\'o}w, 
Pigonia 1, PL-35310 Rzesz{\'o}w, Poland}

\begin{abstract}
We evaluate the cross section for diffractive bremsstrahlung
of a single photon
in the $pp \to pp \gamma$ reaction at high energies and
at forward photon rapidities.
Several differential distributions,
for instance, in ${\rm y}$, $k_{\perp}$ and $\omega$,
the rapidity, the absolute value of the transverse momentum,
and the energy of the photon, respectively, are presented.
We compare the results for our standard approach,
based on QFT and the tensor-pomeron model,
with two versions of soft-photon-approximations, SPA1 and SPA2,
where the radiative amplitudes contain only 
the leading terms proportional to $\omega^{-1}$.
The SPA1, which does not have the correct energy-momentum relations,
performs surprisingly well in the kinematic range considered.
We discuss also azimuthal correlations between outgoing particles.
The azimuthal distributions are not isotropic and
are different for our standard model and SPAs.
We discuss also the possibility of a measurement 
of two-photon-bremsstrahlung 
in the $pp \to pp \gamma \gamma$ reaction.
In our calculations we impose a cut on the relative energy loss
($0.02 < \xi_{i} < 0.1$, $i = 1,2$) of the protons 
where measurements by
the ATLAS Forward Proton (AFP) detectors
are possible.
The AFP requirement for both diffractively scattered protons 
and one forward photon (measured at LHCf) reduces 
the cross section for $p p \to p p \gamma$ almost to zero. 
On the other hand, 
much less cross-section reduction 
occurs for $pp \to pp \gamma \gamma$ when photons are emitted
in opposite sides of the ATLAS interaction point 
and can be measured by two different arms of LHCf.
For the SPA1 ansatz we find 
$\sigma(pp \to pp \gamma \gamma) \simeq 0.03$~nb
at $\sqrt{s} = 13$ TeV
and with the cuts
$0.02 < \xi_{i} < 0.1$, 
$8.5 < {\rm y}_{3} < 9$, $-9 < {\rm y}_{4}< -8.5$.
Our predictions can be verified by ATLAS and LHCf 
combined experiments.
We discuss also the role of the $p p \to p p \pi^0$ background for
single photon production.
\end{abstract}


\maketitle

\section{Introduction}
\label{sec:1}

Bremsstrahlung of photons in nucleon-nucleon collisions
is one of the basic processes in physics.
It was extensively studied in the $pp \to pp \gamma$ reaction
at relatively small c.m. energies where the meson exchanges
are responsible for the underlying $pp$ interaction;
see e.g. \cite{Scholten:2002jb,Haberzettl:2010cg}.
In \cite{Scholten:2002jb} also the two-photon bremsstrahlung 
in $pp$ scattering was considered.
The virtual-photon bremsstrahlung 
in the reactions $NN \to NN (\gamma^{*} \to \ell^{+} \ell^{-})$
was discussed in \cite{Korchin:1995ys,Korchin:1996up}.

At high energies, the $pp \to pp \gamma$ reaction
has not yet been measured.
However, some feasibility studies 
of the measurement of the exclusive diffractive bremsstrahlung cross sections
were performed for RHIC energies 
\cite{Chwastowski:2015mua} and for LHC energies 
using the ATLAS forward detectors
\cite{Chwastowski:2016jkl,Chwastowski:2016zzl}.

In contrast, the inclusive differential production cross-section 
of forward photons in $pp$ collisions was measured 
at $\sqrt{s} = 510$~GeV with the RHICf detector 
(see \cite{Adriani:2022csq})
and at $\sqrt{s} = 0.9, 7$ and 13~TeV
with the LHCf detector 
(see \cite{LHCf:2012stt,LHCf:2011hln,LHCf:2017fnw}).
The LHCf experiment is designed to measure the photons
emitted in the very forward rapidity region 
$|{\rm y}| > 8.4$.
In the ATLAS-LHCf combined analysis \cite{ATLAS:2017rme}
the forward-photon spectra are measured by the LHCf detector,
while the ATLAS inner tracker system
is used to suppress non-diffractive events.\footnote{In this method, the preliminary
photon energy spectrum has been obtained
in two regions of photon rapidity
($8.81 < {\rm y} < 8.99$ and ${\rm y} > 10.94$),
for events with no charged particles having 
$p_{t} > 100$~MeV and $|\eta| < 2.5$,
where $\eta$ is the pseudorapidity.}
In addition, several joint analyses with ATLAS-LHCf
are on-going; 
see the discussions in \cite{Tiberio:2022aej,LHCf:2022nbp}.

In this Letter, 
we discuss exclusive diffractive bremsstrahlung 
of one and two photons
in $pp$ collisions 
for the LHC energy $\sqrt{s} = 13$~TeV 
and at very-forward photon rapidities.
We shall work within the tensor-pomeron model
as proposed in \cite{Ewerz:2013kda}
for soft hadronic high-energy reactions.
The theoretical methods which we shall use
in our present analysis were developed
by us in \cite{Lebiedowicz:2021byo,Lebiedowicz:2022nnn}.
In \cite{Lebiedowicz:2021byo} 
we discussed the soft-photon radiation 
in pion-pion scattering.
Our standard, or also called by us ``exact'', results
for diffractive photon-bremsstrahlung 
were compared to various soft-photon approximations (SPAs).
In \cite{Lebiedowicz:2022nnn} we extended these considerations
to the $pp \to pp \gamma$ reaction at $\sqrt{s} = 13$~TeV,
limiting ourselves to $|{\rm y}| < 5$ and 
$1\; {\rm MeV} < k_{\perp} < 100\; {\rm MeV}$. Here, $k_{\perp}$ 
is the absolute value of the photon transverse momentum.
Recently, in \cite{Lebiedowicz:2023mhe} 
we have discussed various
central-exclusive production (CEP) processes
of single photons.
The CEP processes, for instance the photon-pomeron fusion,
do not play an important role
at forward photon rapidities and can be safely neglected there.

It is also worth noting that
exclusive diffractive photon bremsstrahlung
in high-energy $pp$ collisions at forward rapidities
was discussed earlier
in \cite{Khoze:2010jv,Lebiedowicz:2013xlb,Khoze:2017igg}
within somewhat different approaches.
In general, the bremsstrahlung is not limited to photon production.
The bremsstrahlung-type emission of $\omega$ 
and $\pi^{0}$ mesons in high-energy $pp$ collisions
was calculated in \cite{Cisek:2011vt,Lebiedowicz:2013vya}.

According to our knowledge the exclusive diffractive 
photon bremsstrahlung was not yet identified experimentally.
In order to answer the question whether this is possible 
one needs to consider other processes 
that can be misidentified as bremsstrahlung. 
A dedicated study is in order 
but goes beyond the scope of our present article.
One of the processes which is potentially important in this context
is the $p p \to p p \pi^0$ reaction. The decaying neutral pion
is a source of unwanted photons that can hinder 
the identification of bremsstrahlung photons of interest. 
This reaction was studied by two of us some
time ago \cite{Lebiedowicz:2013vya}. 
In the present letter we wish to briefly
discuss the role of this background contribution.

An interesting proposal to study the forward production 
of ``dark photons'' via bremsstrahlung in $pp$ collisions
with the Forward Physics Facility at the High-Luminosity LHC
was discussed recently in \cite{Foroughi-Abari:2021zbm,Feng:2022inv}.

Our Letter is organized as follows.
In the next section we discuss briefly 
the theoretical formalism.
In Sec.~\ref{sec:2a} we give analytic expressions for
radiative amplitudes for the $pp \to pp \gamma$ reaction
for our standard and approximate approaches.
In Sec.~\ref{sec:2b} we discuss 
two-photon bremsstrahlung in
$pp \to pp \gamma \gamma$.
We present our standard-approach results in Sec.~\ref{sec:3},
along with comparisons to SPAs.
Section~\ref{sec:4} contains our conclusions.

\section{Theoretical formalism}
\label{sec:2}
\subsection{$pp \to pp \gamma$}
\label{sec:2a}
We consider the reaction
\begin{eqnarray}
p (p_{a},\lambda_{a}) + p (p_{b},\lambda_{b}) \to 
p (p_{1}',\lambda_{1}) + p (p_{2}',\lambda_{2}) + \gamma(k, \epsilon)
\label{pp_ppgam}
\end{eqnarray}
at high energies and small momentum transfers.
The momenta are indicated in brackets,
the helicities of the protons are denoted by
$\lambda_{a}, \lambda_{b}, \lambda_{1}, \lambda_{2} 
\in \{1/2, -1/2 \}$,
and $\epsilon$ is the polarization vector of the photon.
The energy-momentum conservation in (\ref{pp_ppgam}) requires
\begin{eqnarray}
p_{a} + p_{b} = p_{1}' + p_{2}' + k\,.
\label{2.4a}
\end{eqnarray}

The kinematic variables are
\begin{eqnarray}
&& s = (p_{a} + p_{b})^{2} = (p_{1}' + p_{2}' + k)^{2}\,, \nonumber \\
&& q_{1} = p_{a} - p_{1}' \,, 
\quad t_{1} = q_{1}^{2}\,, \nonumber \\
&& q_{2} = p_{b} - p_{2}' \,,
\quad t_{2} = q_{2}^{2}\,, \nonumber \\
&& s_{1} = W_{1}^{2} = (p_{1}' + k)^{2} = (p_{a} + q_{2})^{2}\,,
\nonumber \\
&& s_{2} = W_{2}^{2} = (p_{2}' + k)^{2} = (p_{b} + q_{1})^{2}\,.
\label{2.17}
\end{eqnarray}
In the following we work in the overall c.m. system 
where we choose the 3 axis in the direction of $\bpa$ (the beam direction).
The rapidity of the photon is then
\begin{eqnarray}
{\rm y} = \frac{1}{2} \ln \frac{k^{0} + k^{3}}{k^{0} - k^{3}} 
= \tanh^{-1} \left( \frac{k^{3}}{k^{0}} \right)
= - \ln
\tan \frac{\theta}{2} \; ,
\label{rapidity}
\end{eqnarray}
where $\theta$ is the polar angle of $\bk$, 
$\cos\theta = k^{3}/|\bk|$.
For the energy $k^{0}$ of the photon
we use the notation $\omega$.
We introduce the variables $\xi_{1}$ and $\xi_{2}$ which,
to a very good approximation, describe
the fractional energy losses of the protons 
$p (p_{a})$ and $p (p_{b})$
\begin{eqnarray}
\xi_{1} = \frac{p_{b} \cdot q_{1}}{p_{b} \cdot p_{a}} 
= \frac{p_{a}^{0} - p_{1}'^{0}}{p_{a}^{0}} 
+ {\cal O}\left(\frac{M^{2}}{s} \right)\,, \quad
\xi_{2} = \frac{p_{a} \cdot q_{2}}{p_{a} \cdot p_{b}} 
= \frac{p_{b}^{0} - p_{2}'^{0}}{p_{b}^{0}} 
+ {\cal O}\left(\frac{M^{2}}{s} \right)\,.
\label{2.18}
\end{eqnarray}
Here the energies of the incoming and outgoing protons, respectively, are
\begin{eqnarray}
p_{a}^{0} &=& p_{b}^{0} = \frac{\sqrt{s}}{2}  \,,\\
p_{1,2}'^{0} &=& 
\frac{1}{2 \sqrt{s}} (s + m_{p}^{2} - s_{2,1})\,,
\label{2.18_aux}
\end{eqnarray}
and we set 
$M^{2} = {\rm max}(m_{p}^{2},|t_{1}|,|t_{2}|, k_{\perp}^{2})$.
Alternatively, the proton relative energy-loss parameters
can be expressed by the kinematical variables of the photon,
\begin{eqnarray}
\xi_{1} = \frac{k_{\perp}}{\sqrt{s}} \exp({\rm y})
+ {\cal O}\left(\frac{M^{2}}{s} \right)\,, \quad
\xi_{2} = \frac{k_{\perp}}{\sqrt{s}} \exp(-{\rm y})
+ {\cal O}\left(\frac{M^{2}}{s} \right)\,.
\label{2.19}
\end{eqnarray}

The cross section for the photon yield can be calculated as follows
\begin{eqnarray}
d\sigma({pp \to pp \gamma}) &=&
\frac{1}{2\sqrt{s(s-4 m_{p}^{2})}}
\frac{d^{3}k}{(2 \pi)^{3} \,2 k^{0}}
\int 
\frac{d^{3}p_{1}'}{(2 \pi)^{3} \,2 p_{1}'^{0}}
\frac{d^{3}p_{2}'}{(2 \pi)^{3} \,2 p_{2}'^{0}}
\nonumber \\
&&\times 
(2 \pi)^{4} \delta^{(4)}(p_{1}'+p_{2}'+k-p_{a}-p_{b})
\frac{1}{4}
\sum_{p\; {\rm spins}}
{\cal M}_{\mu} {\cal M}_{\nu}^{*} (-g^{\mu \nu})\,; \quad 
\label{xs_2to3}
\end{eqnarray}
see Eqs.~(2.33)--(2.35) of \cite{Lebiedowicz:2022nnn}.
${\cal M}_{\mu}$ is the radiative amplitude.

Our standard photon-bremsstrahlung amplitude,
${\cal M}_{\mu}^{\rm standard}$,
treated in the tensor-pomeron approach,
see (2.62) and (B3) of \cite{Lebiedowicz:2022nnn},
includes 6 diagrams shown in Fig.~3(a)--(f) of \cite{Lebiedowicz:2022nnn}. 
The amplitudes (a), (b), (d), and (e), 
corresponding to photon emission 
from the external protons, are determined 
by the off-shell $pp$ elastic scattering amplitude. 
The contact terms, (c) and (f), are needed 
in order to satisfy gauge-invariance constraints.
For details how to calculate these standard results
we refer the reader to Sec.~II~C 
and Appendix~B of \cite{Lebiedowicz:2022nnn}.

In the following, we shall compare our standard results
to two soft-photon approximations, SPA1 and SPA2,
as defined in Sec.~III of \cite{Lebiedowicz:2022nnn}.
In both SPAs we keep only the pole terms $\propto \omega^{-1}$.
We consider only the pomeron-exchange contribution for the radiative 
amplitudes, the leading term at high energies.

In SPA1, the radiative amplitude has the form
\begin{eqnarray}
{\cal M}_{\mu}
\to {\cal M}_{\mu, \;{\rm SPA1}}
= e{\cal M}^{({\rm on\; shell})\,pp}(s,t)
\Big[ 
-\frac{p_{a \mu}}{(p_{a} \cdot k)}
+\frac{p_{1 \mu}}{(p_{1} \cdot k)}
-\frac{p_{b \mu}}{(p_{b} \cdot k)}
+\frac{p_{2 \mu}}{(p_{2} \cdot k)} \Big], \qquad
\label{SPA1}
\end{eqnarray}
where ${\cal M}^{({\rm on\; shell})\,pp}(s,t)$ 
is the amplitude for on-shell $pp$-scattering 
\begin{eqnarray}
&&p (p_{a},\lambda_{a}) + p (p_{b},\lambda_{b}) \to 
p (p_{1},\lambda_{1}) + p (p_{2},\lambda_{2}) \,, \nonumber \\
&&p_{a} + p_{b} = p_{1} + p_{2}\,;
\label{pp_pp_on_shell}
\end{eqnarray}
see (2.19) and (3.1) of \cite{Lebiedowicz:2022nnn}.
The inclusive photon cross section for the SPA1 case is
\begin{eqnarray}
d\sigma({pp \to pp \gamma})_{\,\rm SPA1} &=&
\frac{d^{3}k}{(2 \pi)^{3} \,2 k^{0}}
\int d^{3}p_{1} \,d^{3}p_{2}\,e^{2}\;
\frac{d\sigma(pp \to pp)}{d^{3}p_{1}d^{3}p_{2}}\nonumber \\
&&\times 
\Big[ 
-\frac{p_{a \mu}}{(p_{a} \cdot k)}
+\frac{p_{1 \mu}}{(p_{1} \cdot k)}
-\frac{p_{b \mu}}{(p_{b} \cdot k)}
+\frac{p_{2 \mu}}{(p_{2} \cdot k)} \Big]\nonumber \\
&&\times 
\Big[ 
-\frac{p_{a \nu}}{(p_{a} \cdot k)}
+\frac{p_{1 \nu}}{(p_{1} \cdot k)}
-\frac{p_{b \nu}}{(p_{b} \cdot k)}
+\frac{p_{2 \nu}}{(p_{2} \cdot k)} \Big] (-g^{\mu \nu})\,.
\quad
\label{SPA1_xs}
\end{eqnarray}
Here
\begin{eqnarray}
\frac{d\sigma(pp \to pp)}{d^{3}p_{1}d^{3}p_{2}} &=&
\frac{1}{2\sqrt{s(s-4 m_{p}^{2})}}\,
\frac{1}{(2 \pi)^{3} \,2 p_{1}^{0}\,(2 \pi)^{3} \,2 p_{2}^{0}} 
\nonumber\\
&&\times (2 \pi)^{4} \delta^{(4)}(p_{1}+p_{2}-p_{a}-p_{b})\,
\frac{1}{4}\sum_{p \, \rm spins}
|{\cal M}^{({\rm on\; shell})\,pp}(s,t)|^{2}\,,\qquad \;
\label{SPA1_xs_aux}
\end{eqnarray}
where we neglect the photon momentum $k$
in the energy-momentum conserving $\delta^{(4)}(.)$ function.
For SPA1 results we impose restrictions on 
the proton's relative energy loss variables $\xi_{i}$ by using (\ref{2.19})
neglecting terms of ${\cal O}(M^{2}/s)$.

In the SPA2 case, we keep the exact energy-momentum relation (\ref{2.4a}).
Here we calculate the photon yield using (\ref{xs_2to3}) 
replacing the radiative amplitude as follows
\begin{eqnarray}
{\cal M}_{\mu}
\to  {\cal M}_{\mu, \;{\rm SPA2}}
= {\cal M}_{{\Pom},\mu}^{(a + b + c)\,1}
+
{\cal M}_{{\Pom},\mu}^{(d + e + f)\,1}\,.
\label{SPA2}
\end{eqnarray}
The explicit expressions of these terms
are given by (3.4), (B4), and (B15) of \cite{Lebiedowicz:2022nnn}. 

\subsection{$pp \to pp \gamma \gamma$}
\label{sec:2b}

Here we consider the reaction
\begin{eqnarray}
p (p_{a}) + p (p_{b}) \to 
p (p_{1}') + p (p_{2}') + \gamma(k_{3})+ \gamma(k_{4})\,.
\label{pp_ppgamgam}
\end{eqnarray}
We shall study this reaction
under specific conditions. 
We shall require that one photon is emitted 
at forward and one at backward rapidities,
\mbox{$8.5 < {\rm y}_{3} < 9$} and $-9 < {\rm y}_{4} < -8.5$,
respectively,
and that $0.02 < \xi_{1,2} < 0.1$.~\footnote{The AFP acceptance 
is limited by $\xi$ and transverse momentum $p_{t,p}$
of the scattered protons;
see Fig.~8 of \cite{Trzebinski:2014vha}.}

For the calculation of the radiative amplitudes we use SPA1.
Here in the $2 \to 4$ kinematics for SPA1 we define
\begin{eqnarray}
&&\xi_{1} = \frac{k_{\perp 3}}{\sqrt{s}} \exp({\rm y}_{3})
        + \frac{k_{\perp 4}}{\sqrt{s}} \exp({\rm y}_{4})\,, \nonumber \\
&&\xi_{2} = \frac{k_{\perp 3}}{\sqrt{s}} \exp(-{\rm y}_{3})
        + \frac{k_{\perp 4}}{\sqrt{s}} \exp(-{\rm y}_{4})\,.
\label{xi_2to4}
\end{eqnarray}
%

\begin{figure}[!h]
\includegraphics[width=9.5cm]{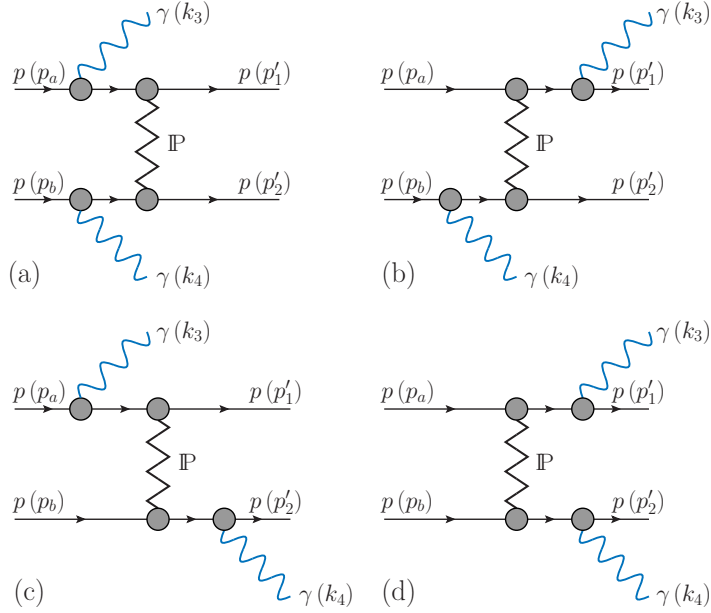}
\caption{Diffractive two-photon bremsstrahlung diagrams 
for the reaction $pp \to pp \gamma \gamma$ (\ref{pp_ppgamgam}) 
with exchange of the pomeron $\Pom$.
These four diagrams (a -- d) contribute to the SPA1 amplitude
(\ref{SPA1_gamgam}).
In addition there are the diagrams where 
the two photons are emitted
from the $p_{a}$-$p_{1}'$ line 
or from the $p_{b}$-$p_{2}'$ line,
and various contact terms.
These diagrams are not shown here.}
\label{fig:pp_brems_gamgam_SPA1}
\end{figure}
We shall see below in Sec.~\ref{sec:3} that, indeed, 
the above cuts on $\xi_{1}$ and $\xi_{2}$
assure that to a good approximation we can restrict ourselves
to the diagrams
shown in Fig.~\ref{fig:pp_brems_gamgam_SPA1}.
Then, the two-photon bremsstrahlung amplitude has the form
\begin{eqnarray}
{\cal M}_{\mu \nu, \;{\rm SPA1}}
= e^{2} {\cal M}^{({\rm on\; shell})\,pp}(s,t)
\Big[ 
-\frac{p_{a \mu}}{(p_{a} \cdot k_{3})}
+\frac{p_{1 \mu}}{(p_{1} \cdot k_{3})} \Big]
\Big[ 
-\frac{p_{b \nu}}{(p_{b} \cdot k_{4})}
+\frac{p_{2 \nu}}{(p_{2} \cdot k_{4})} \Big], \qquad \quad
\label{SPA1_gamgam}
\end{eqnarray}
and the inclusive two-photon cross section is
\begin{eqnarray}
d\sigma({pp \to pp \gamma \gamma})_{\,\rm SPA1} &=&
\frac{d^{3}k_{3}}{(2 \pi)^{3} \,2 k_{3}^{0}}
\frac{d^{3}k_{4}}{(2 \pi)^{3} \,2 k_{4}^{0}}
\int d^{3}p_{1} \,d^{3}p_{2}\,e^{4}\;
\frac{d\sigma(pp \to pp)}{d^{3}p_{1}d^{3}p_{2}}\nonumber \\
&&\times 
\Big[ 
-\frac{p_{a \mu}}{(p_{a} \cdot k_{3})}
+\frac{p_{1 \mu}}{(p_{1} \cdot k_{3})} \Big]
\Big[ 
-\frac{p_{b \nu}}{(p_{b} \cdot k_{4})}
+\frac{p_{2 \nu}}{(p_{2} \cdot k_{4})} \Big] \nonumber \\
&&\times 
\Big[ 
-\frac{p_{a \alpha}}{(p_{a} \cdot k_{3})}
+\frac{p_{1 \alpha}}{(p_{1} \cdot k_{3})} \Big]
\Big[ 
-\frac{p_{b \beta}}{(p_{b} \cdot k_{4})}
+\frac{p_{2 \beta}}{(p_{2} \cdot k_{4})} \Big]  
\nonumber \\
&&\times 
(-g^{\mu \alpha}) (-g^{\nu \beta}) \,.
\quad
\label{SPA1_xs_gamgam}
\end{eqnarray}
%

\section{Results}
\label{sec:3}

\subsection{Single photon emission}
\label{sec:single}

In the following we consider explicitly only photon emission
in very forward direction (${\rm y} \gg 1$), 
where the photon is emitted predominantly from
$p(p_{a})$ plus $p(p_{1}')$; see (\ref{pp_ppgam}).
As we see from (\ref{2.19}) we have here
$\xi_{1}$ sizeable but $\xi_{2}$ very small.
In fact, then the proton $p(p_{2}')$, 
having nearly the same energy as the incoming proton $p(p_{b})$,
cannot be measured by the present AFP detectors.
Therefore, requiring for single emission that both final state protons
are measured by the AFP detectors reduces the cross section
essentially to zero. 
Thus, for photon emission at ${\rm y} \gg 1$
we consider only detection of $p(p_{1}')$ in the AFP detector.
Of course, for ${\rm y} \ll -1$ the roles of $p(p_{1}')$ and $p(p_{2}')$
are interchanged and all distributions
shown below for ${\rm y} \gg 1$ are easily transferred to ${\rm y} \ll -1$.

\begin{figure}[!ht]
(a)\includegraphics[width=0.44\textwidth]{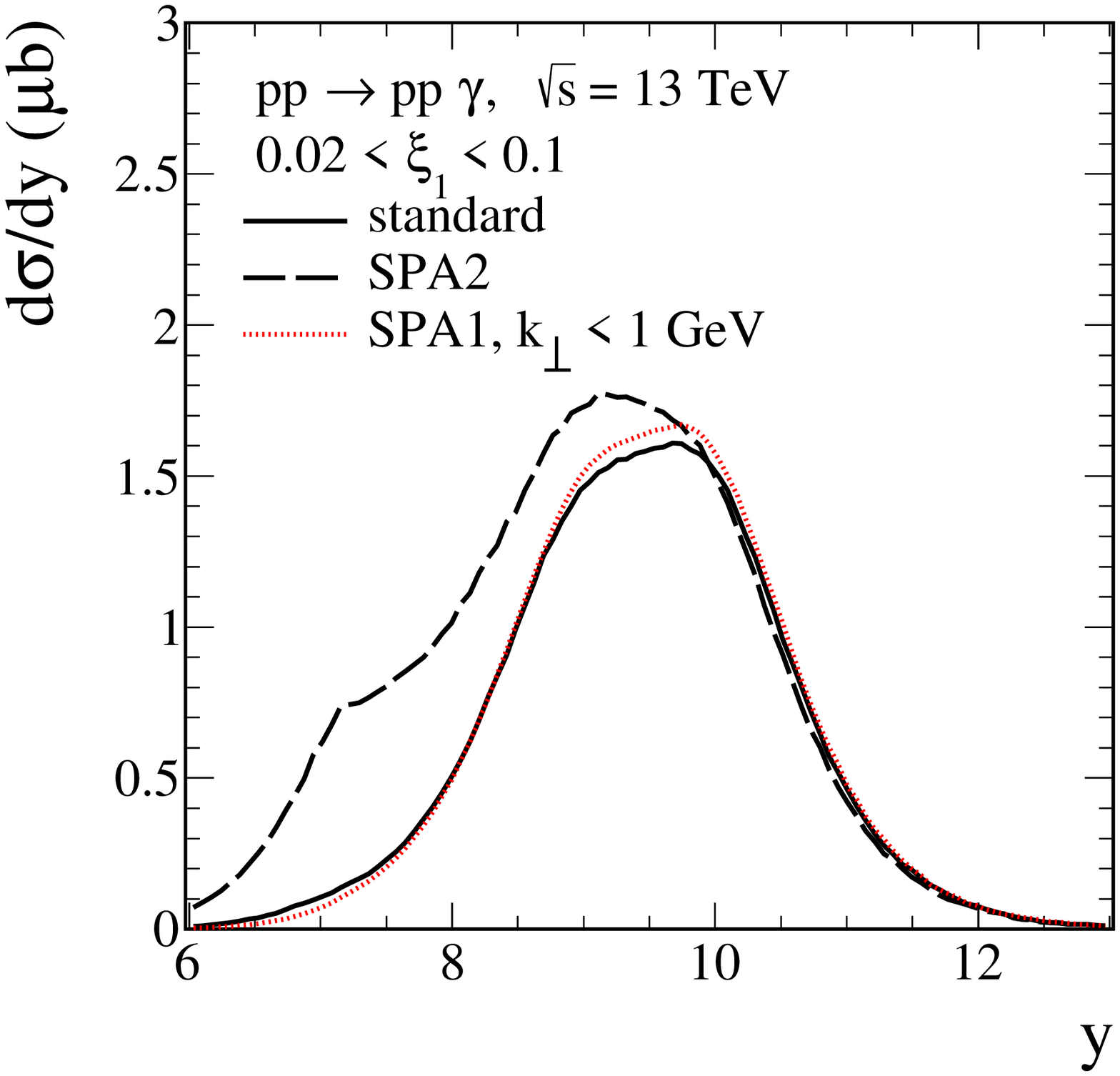}
(b)\includegraphics[width=0.44\textwidth]{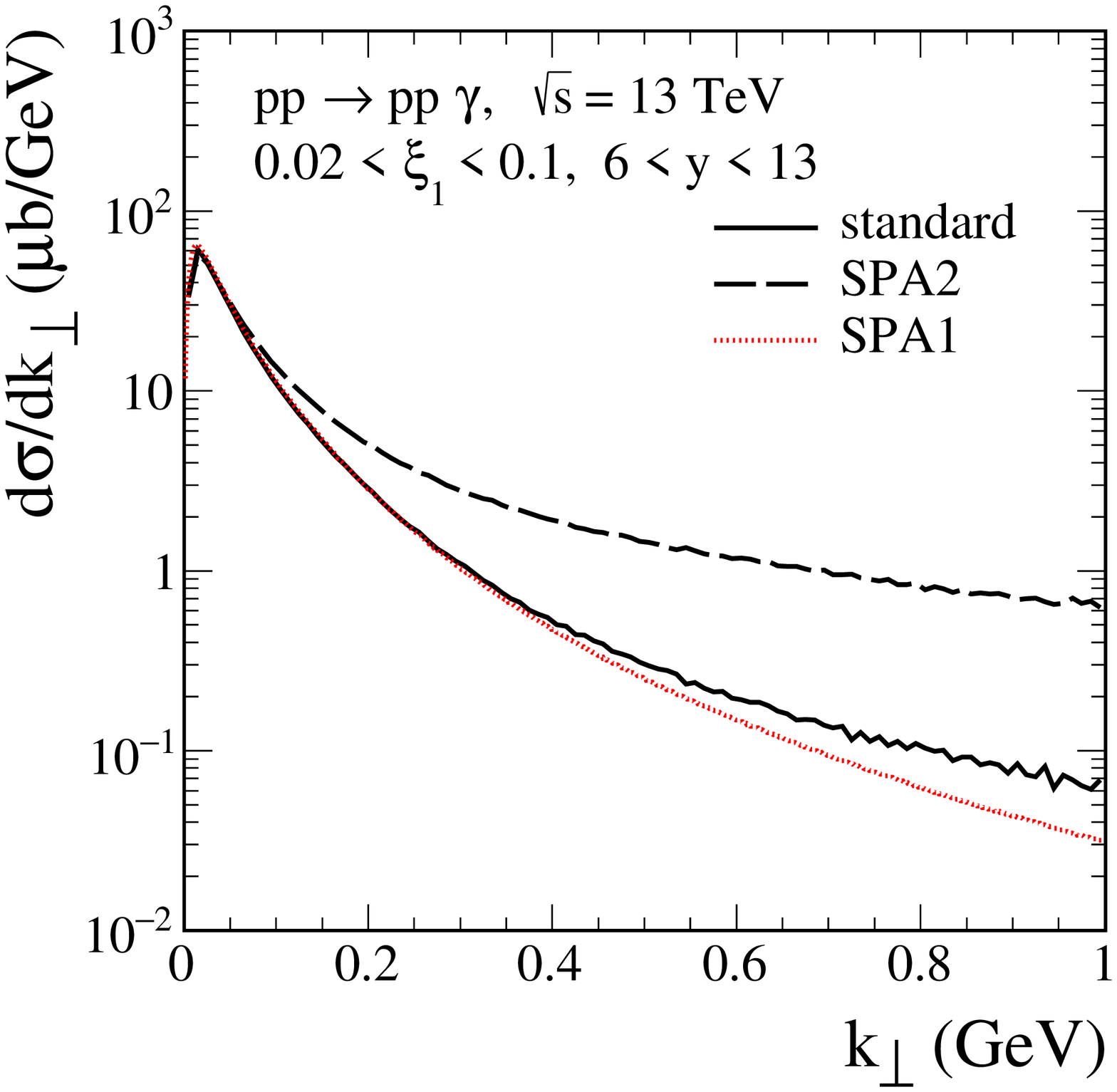}
(c)\includegraphics[width=0.44\textwidth]{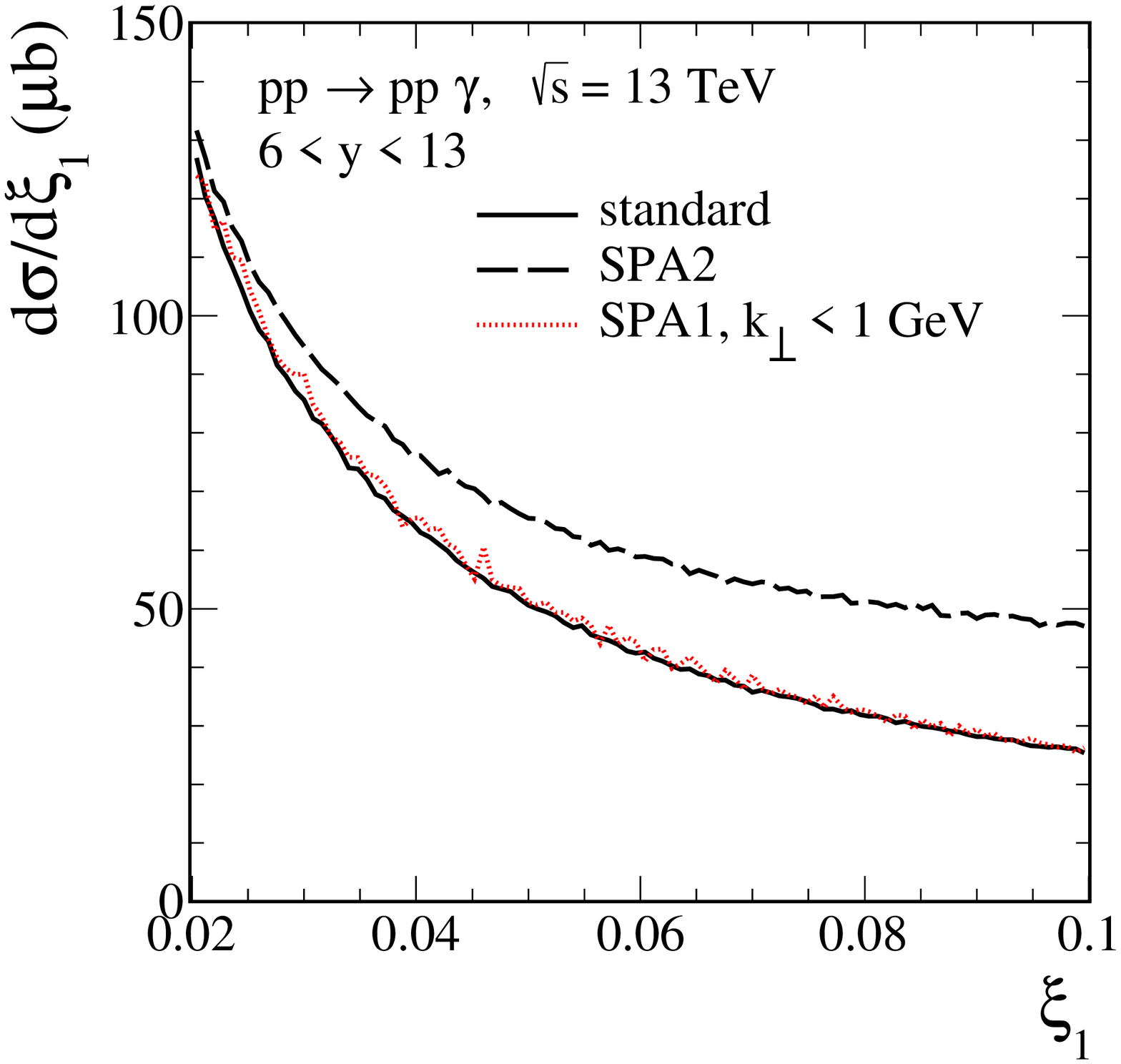}
(d)\includegraphics[width=0.44\textwidth]{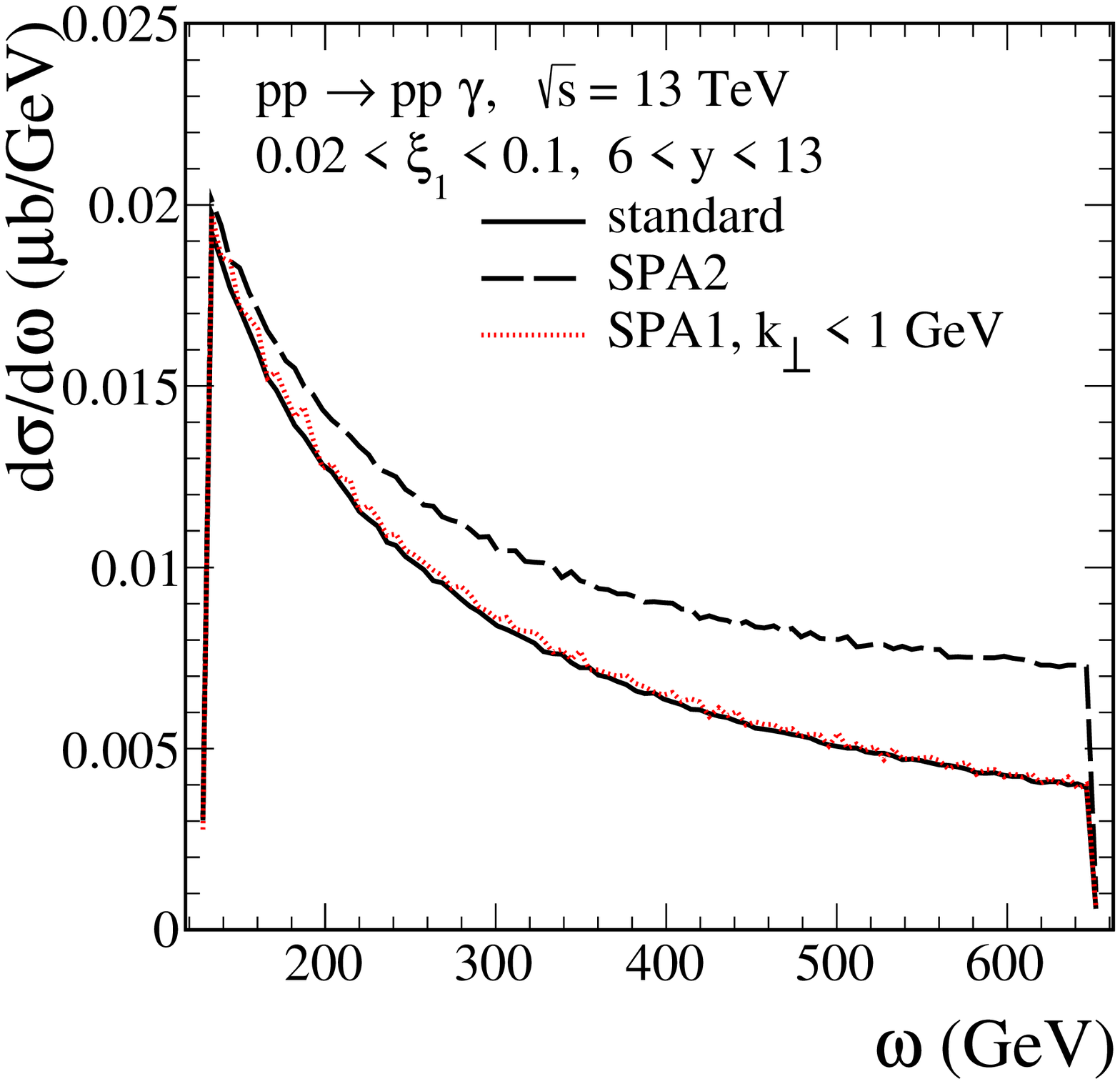}
\caption{\label{fig:1}
\small
The differential distributions 
in the rapidity of the photon (a), 
in the transverse momentum of the photon (b), 
in the energy-loss variable $\xi_{1}$ (c), and 
in the energy of the photon (d)
for the $pp \to pp \gamma$ reaction
via bremsstrahlung.
The calculations were done for $\sqrt{s} = 13$~TeV, $6 < {\rm y} < 13$,
and with the cut $0.02 < \xi_{1} < 0.1$.
For SPA1 an additional cut $k_{\perp} < 1$~GeV was imposed.
The solid line corresponds to our standard bremsstrahlung model,
the black long-dashed line corresponds to SPA2 (\ref{SPA2}), and 
the red dotted line corresponds to SPA1 (\ref{SPA1}).
The oscillations in the SPA1 results are of numerical origin}
\end{figure}

In Fig.~\ref{fig:1} we show the distributions
in rapidity of the photon [see the panel (a)], 
in the absolute value of the transverse momentum of the photon 
[see the panel (b)],
in $\xi_{1}$ [see the panel (c)], and 
in the energy of the photon [see the panel (d)].
Shown are the results in the forward rapidity region 
for our standard approach (standard bremsstrahlung results)
for the $pp \to pp \gamma$ reaction 
together with the results obtained via SPA1 and SPA2
discussed in Sec.~\ref{sec:2a}.
We see from the panels (a) and (b) of Fig.~\ref{fig:1}
that bremsstrahlung photons
are emitted predominantly in very forward-rapidity region 
$9 < {\rm y} < 10$
and with small values of $k_{\perp}$;
see also the left panel of Fig.~\ref{fig:map_kty}.
Forward photons will be measured by the LHCf experiment
in the regions $8.5 \lesssim {\rm y} \lesssim 9$ 
and ${\rm y} \gtrsim 11$.
Due to the cut $0.02 < \xi_{1} < 0.1$
the energy of the photons is limited to 
$130~{\rm GeV} < \omega < 650~{\rm GeV}$;
see the panels (c) and (d) of Fig.~\ref{fig:1}.
Note, that due to the cuts specified in the figure legend 
the distributions in $k_{\perp}$ and $\omega$
have no singularity for $k_{\perp} \to 0$,
respectively $\omega \to 0$.
The $k_{\perp}$ distribution
reaches a maximum at $k_{\perp} \sim 0.014$~GeV,
and then it quickly decreases with increasing $k_{\perp}$.

In the SPA1, the photon momentum $k$ was, on purpose,
omitted in the energy-momentum conserving $\delta$ function
in the evaluation of the cross section [see (\ref{SPA1_xs}) and (\ref{SPA1_xs_aux})].
Here, the cross section is integrated
over $k_{\perp}$ from $k_{\perp {\rm min}}$
to a maximal value $k_{\perp {\rm max}}$ which we set to 1~GeV.
In the SPA2, the correct $2 \to 3$ kinematics is used.
Recall that in both SPAs
we keep only the pole terms $\propto \omega^{-1}$
in the radiative amplitudes.
We see from Fig.~\ref{fig:1} that the SPA1,
which does not have the correct energy-momentum relations,
performs surprisingly well in the kinematic range considered.
For the SPA2, the deviations from our standard result
increase rapidly with growing $k_{\perp}$ and $\omega$.
From this comparison we see the importance
of the interference 
between the pole term and the non-leading,
but numerically large,
terms occurring in the radiative amplitudes.
It is essential to add coherently all the various parts of the amplitude
for the bremsstrahlung-type emission of photons
in order not to miss important interference effects.
For more details on the size of various contributions
we refer to the discussions in \cite{Lebiedowicz:2022nnn}
(see, e.g., Fig.~17 there).

\begin{figure}[!ht]
(a)\includegraphics[width=0.44\textwidth]{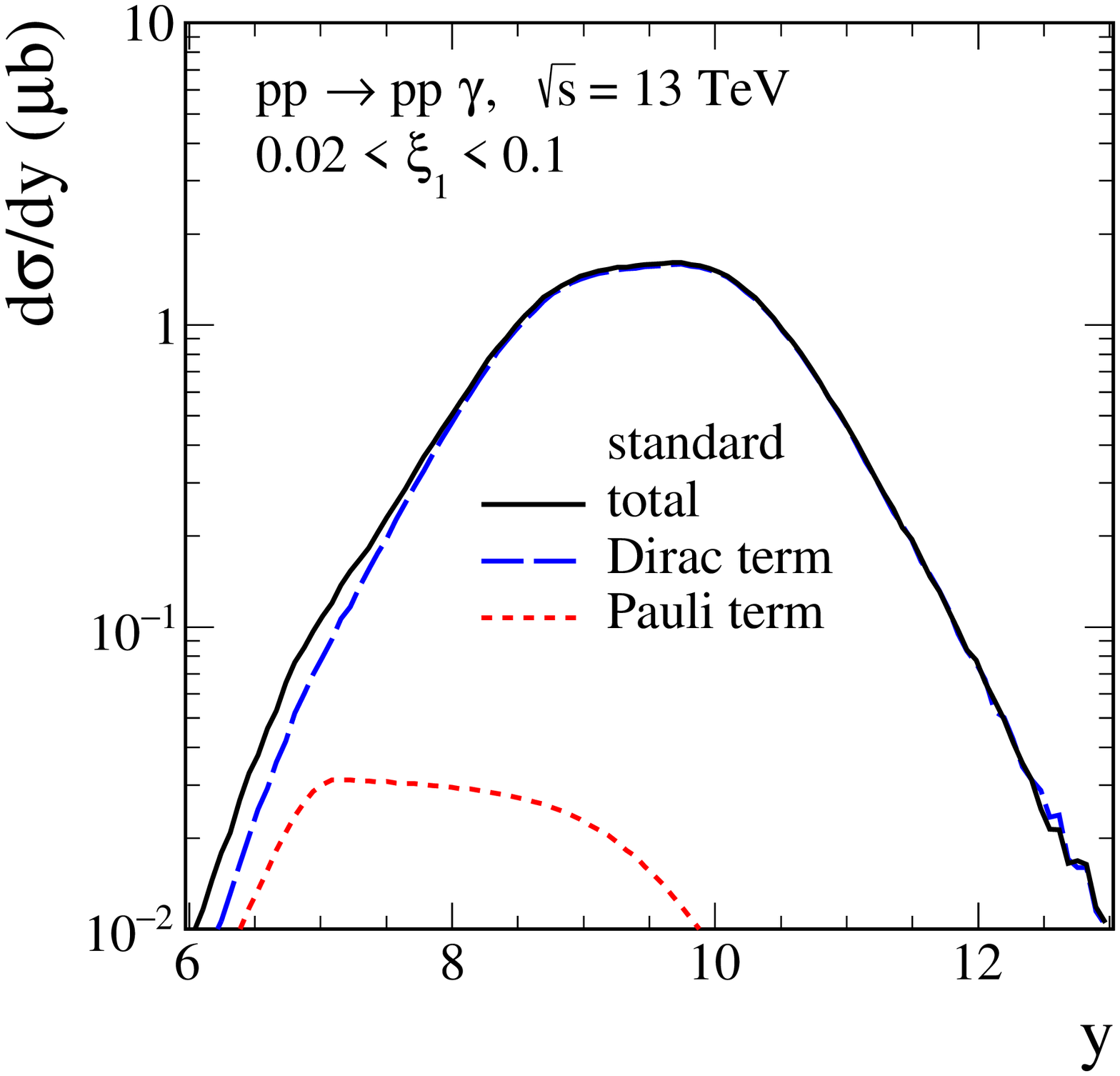}
(b)\includegraphics[width=0.44\textwidth]{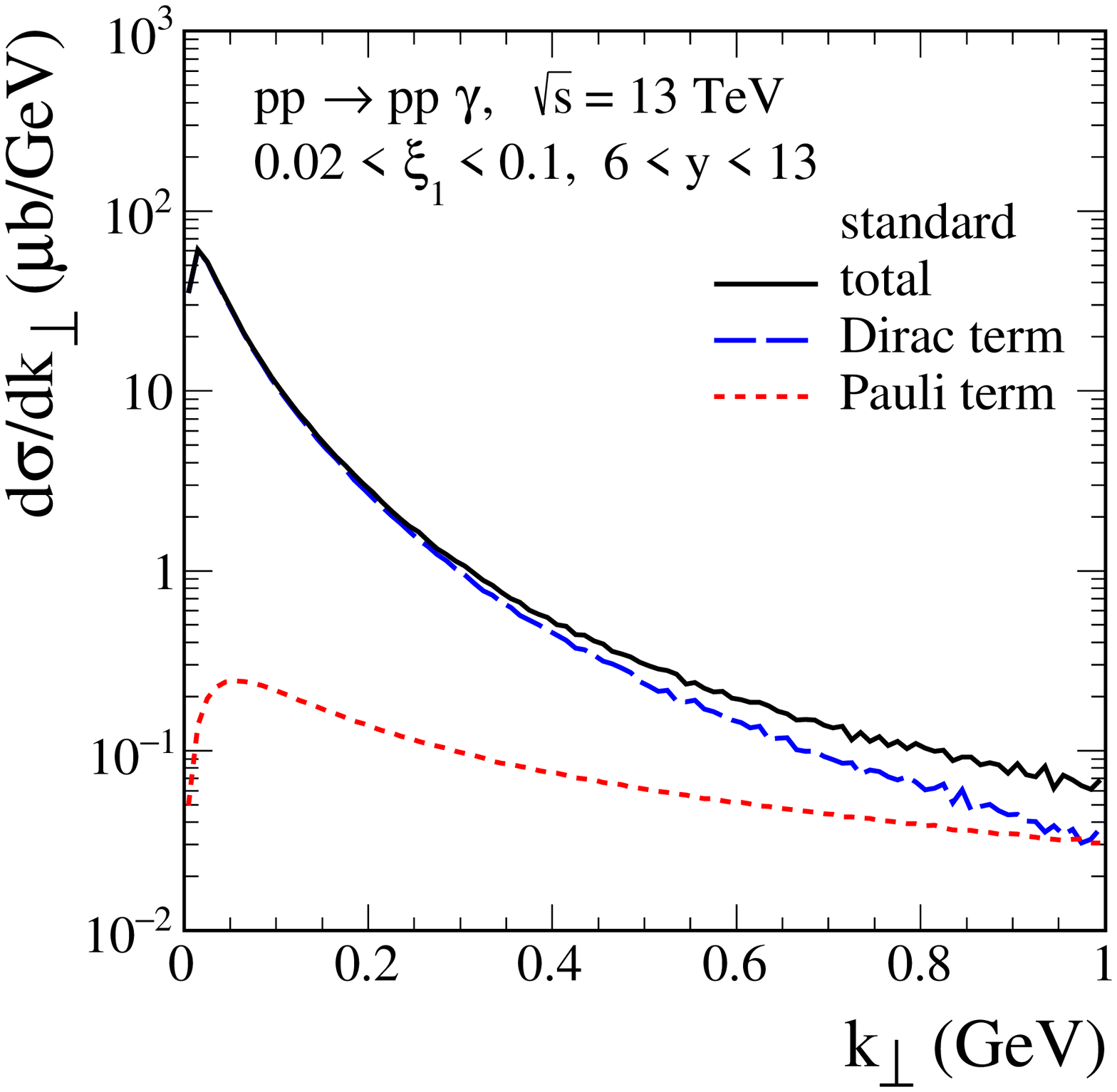}
(c)\includegraphics[width=0.44\textwidth]{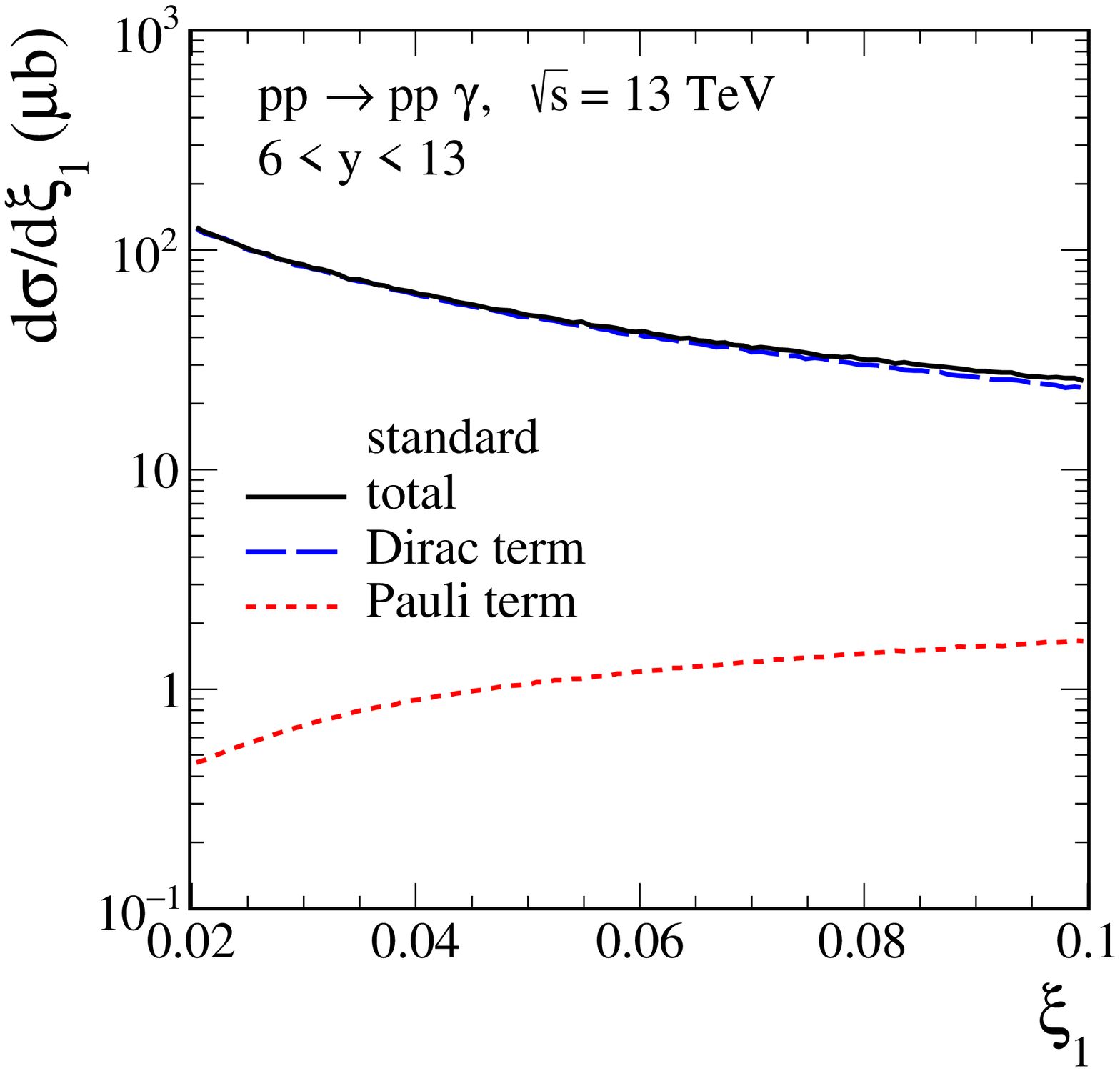}
(d)\includegraphics[width=0.44\textwidth]{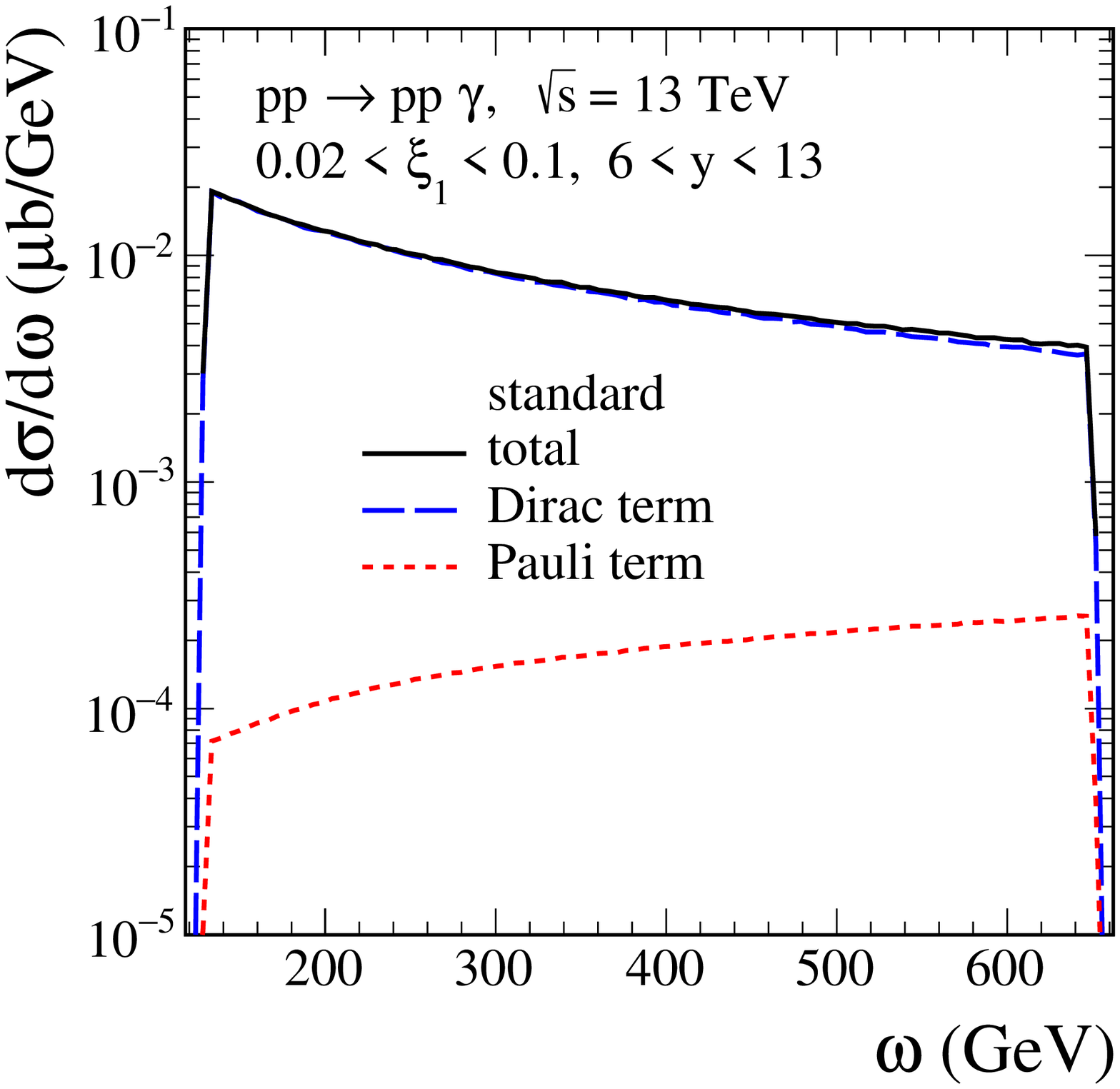}
\caption{\label{fig:1_deco}
\small
The differential cross sections
for the $pp \to pp \gamma$ reaction
via bremsstrahlung as in Fig.~\ref{fig:1}
but here shown are only standard complete results
(total) and the results for the Dirac and Pauli terms alone.}
\end{figure}

\begin{figure}[!ht]
\includegraphics[width=0.49\textwidth]{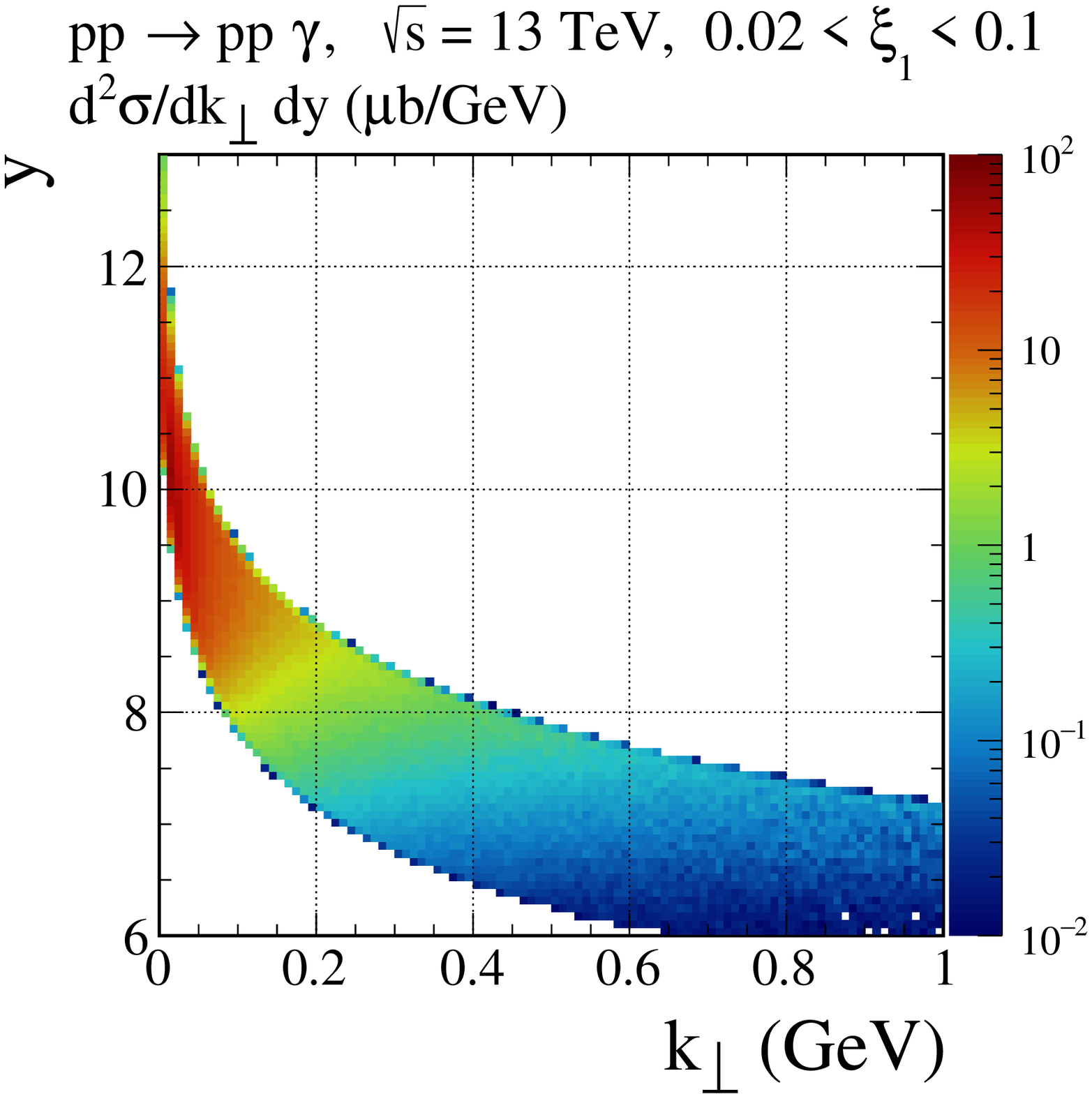}
\includegraphics[width=0.49\textwidth]{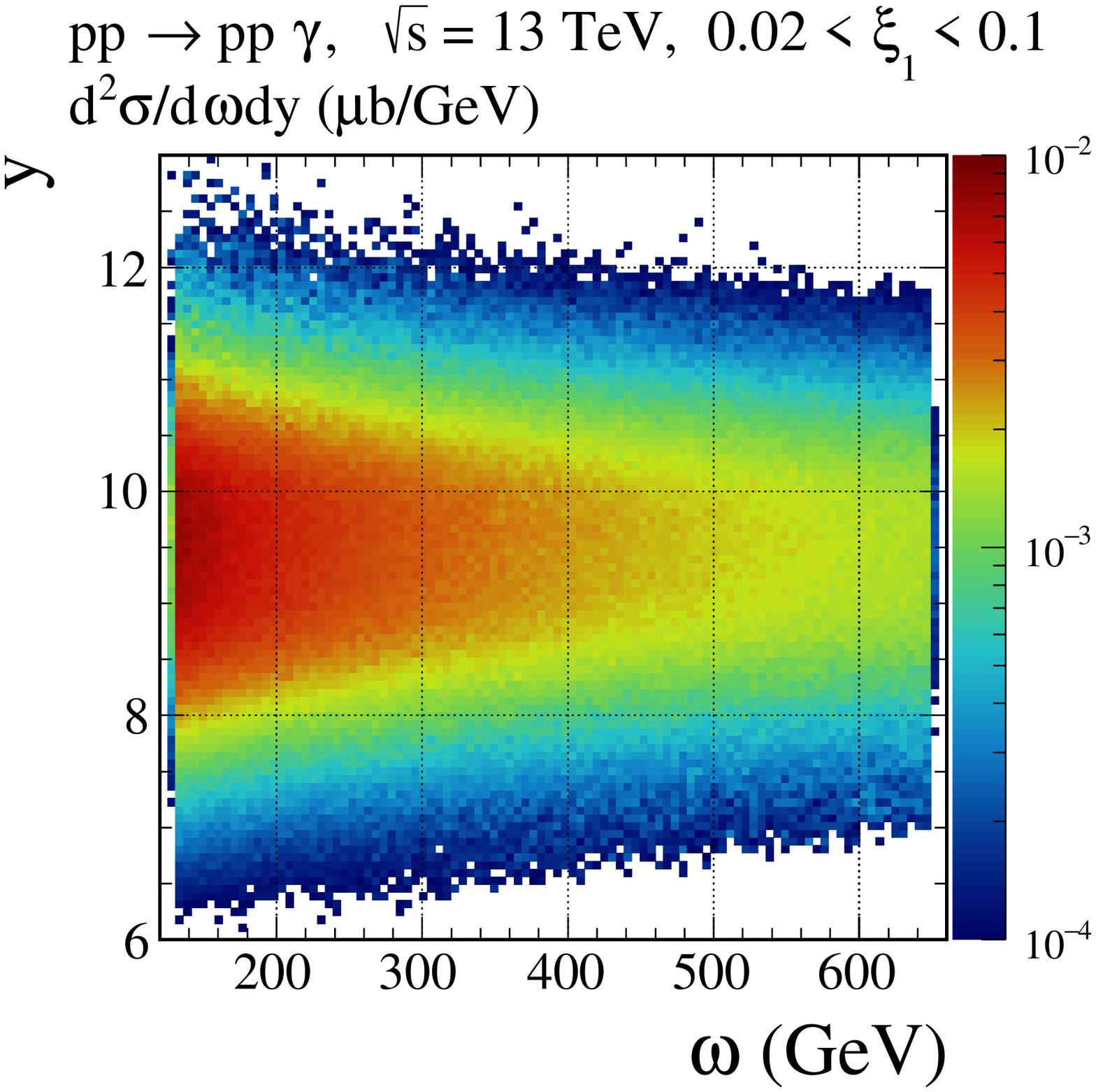}
\caption{\label{fig:map_kty}
\small
The two-dimensional distributions in ($k_{\perp}, {\rm y}$)
and in ($\omega, {\rm y}$) for the $pp \to pp \gamma$ reaction
for our standard bremsstrahlung model
calculated for $\sqrt{s} = 13$~TeV,
$6 < {\rm y} < 13$, and $0.02 < \xi_{1} < 0.1$.}
\end{figure}

In the standard result, all contributions to the radiative amplitude
with Dirac and Pauli terms are included.
Figure~\ref{fig:1_deco} shows the complete (total) standard results
and the results for Dirac and Pauli terms individually.
The anomalous magnetic moment of the proton (Pauli term)
plays an important role for larger values of $k_{\perp}$ and $\omega$.


In Fig.~\ref{fig:map_kty} we show the two-dimensional differential
cross sections in the $k_{\perp}$-${\rm y}$ plane (left panel)
and in the $\omega$-${\rm y}$ plane (right panel)
imposing a typical cut $0.02 < \xi_{1} < 0.1$ on the fractional energy loss
of the 'emitting' proton.

\begin{figure}[!ht]
\includegraphics[width=0.44\textwidth]{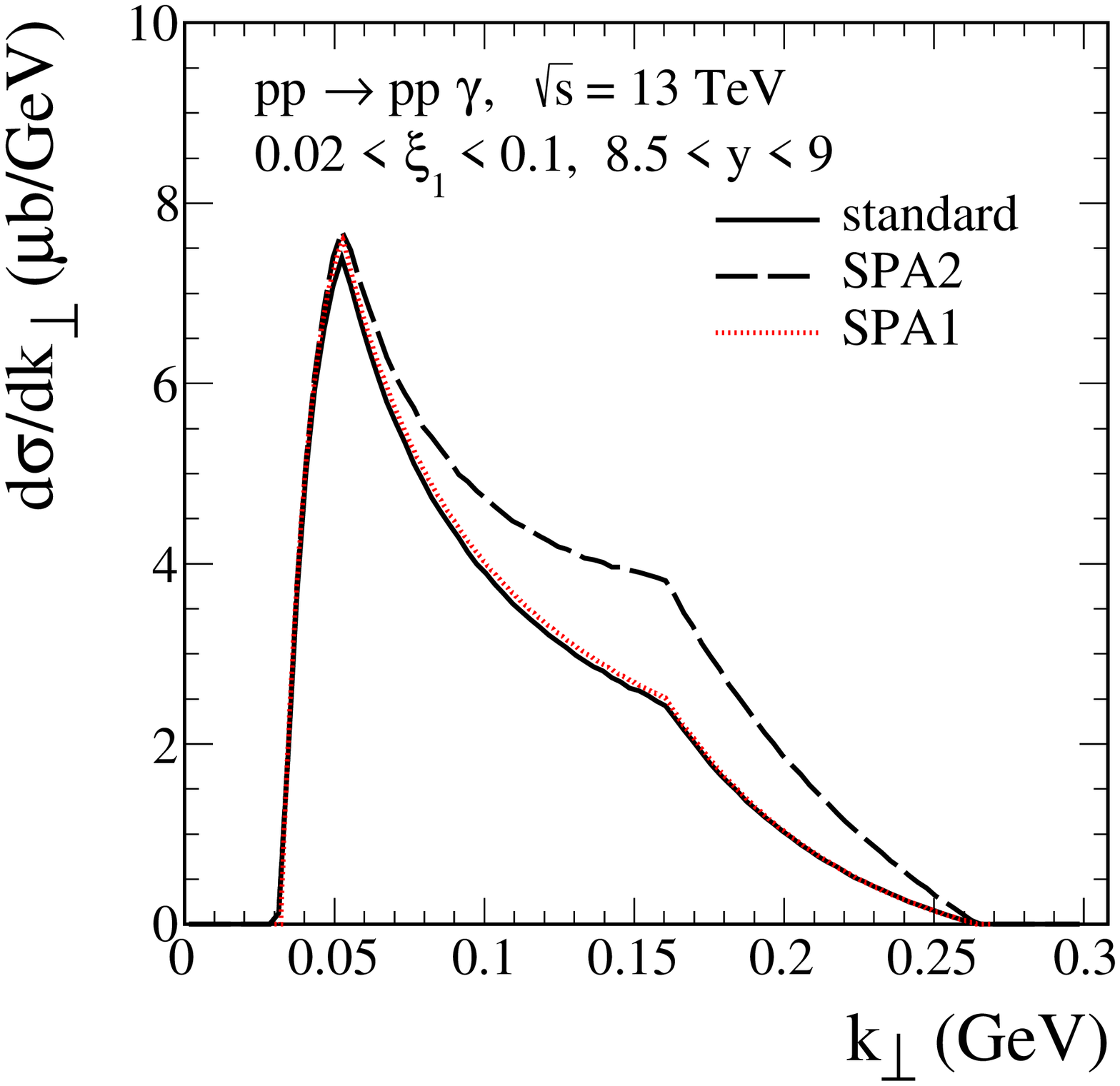}
\includegraphics[width=0.44\textwidth]{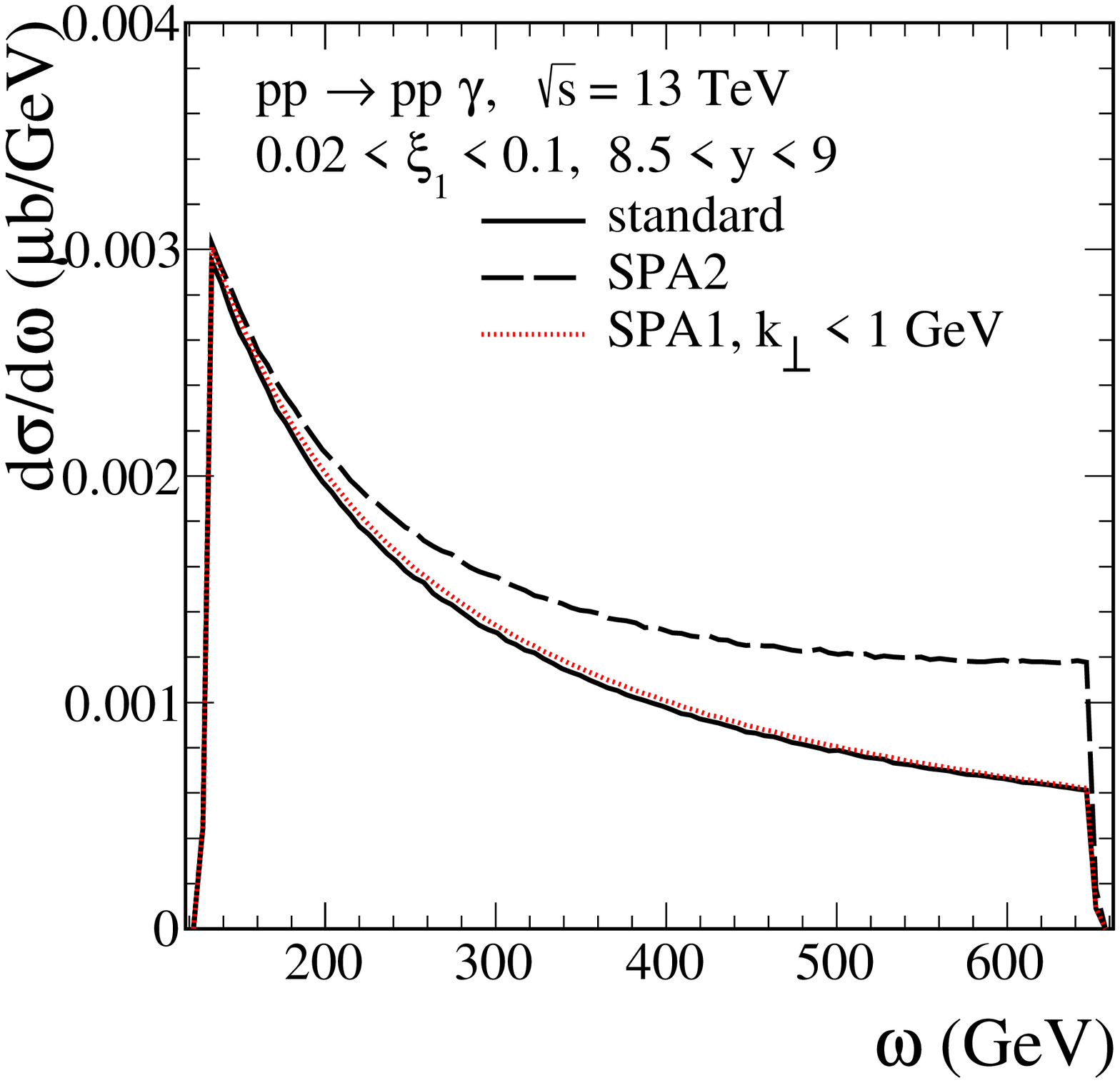}
\caption{\label{fig:1_8.5_9}
\small
The differential distributions 
in the transverse momentum of the photon (left) and 
in the energy of the photon (right)
for the $pp \to pp \gamma$ reaction.
The calculations were done for $\sqrt{s} = 13$~TeV,
$8.5 < {\rm y} < 9$, and $0.02 < \xi_{1} < 0.1$.
The meaning of the lines is the same as in Fig.~\ref{fig:1}.}
\end{figure}

In Fig.~\ref{fig:1_8.5_9} we show the results for
$d\sigma/dk_{\perp}$ and $d\sigma/d\omega$ for $\sqrt{s} = 13$~TeV,
$0.02 < \xi_{1} < 0.1$, 
as in Figs.~\ref{fig:1}(b) and \ref{fig:1}(d) but now for a more restrictive
${\rm y}$ cut, $8.5 < {\rm y} < 9$.
We see that the $k_{\perp}$ and $\omega$
distributions are reduced by a factor of order 10
compared to their counterparts in Figs.~\ref{fig:1}~(b) and (d).
The kink at $k_{\perp} \approx 0.17$~GeV
is due to the cut on ${\rm y}$.

\begin{figure}[!ht]
\includegraphics[width=0.44\textwidth]{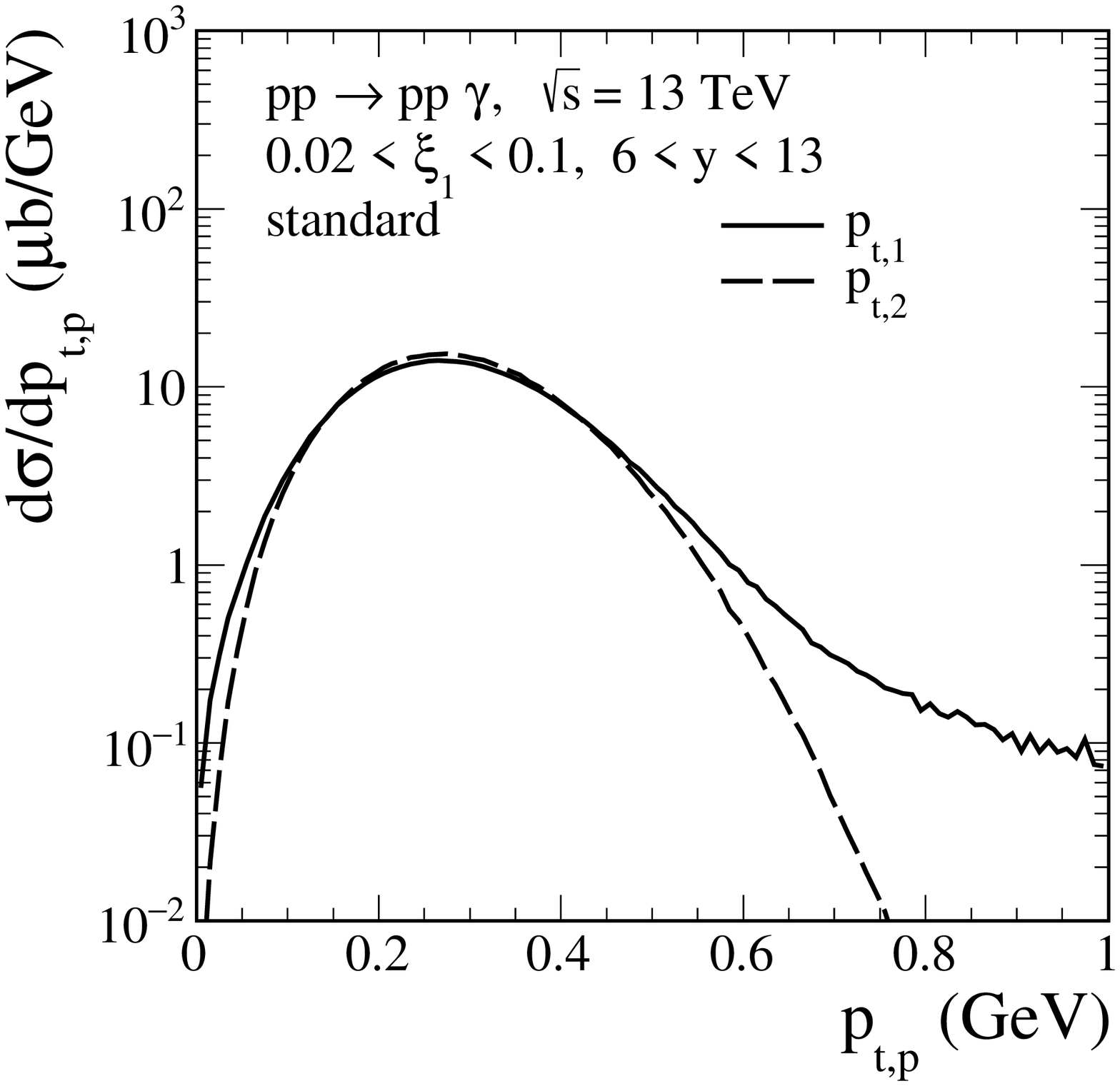}
\includegraphics[width=0.44\textwidth]{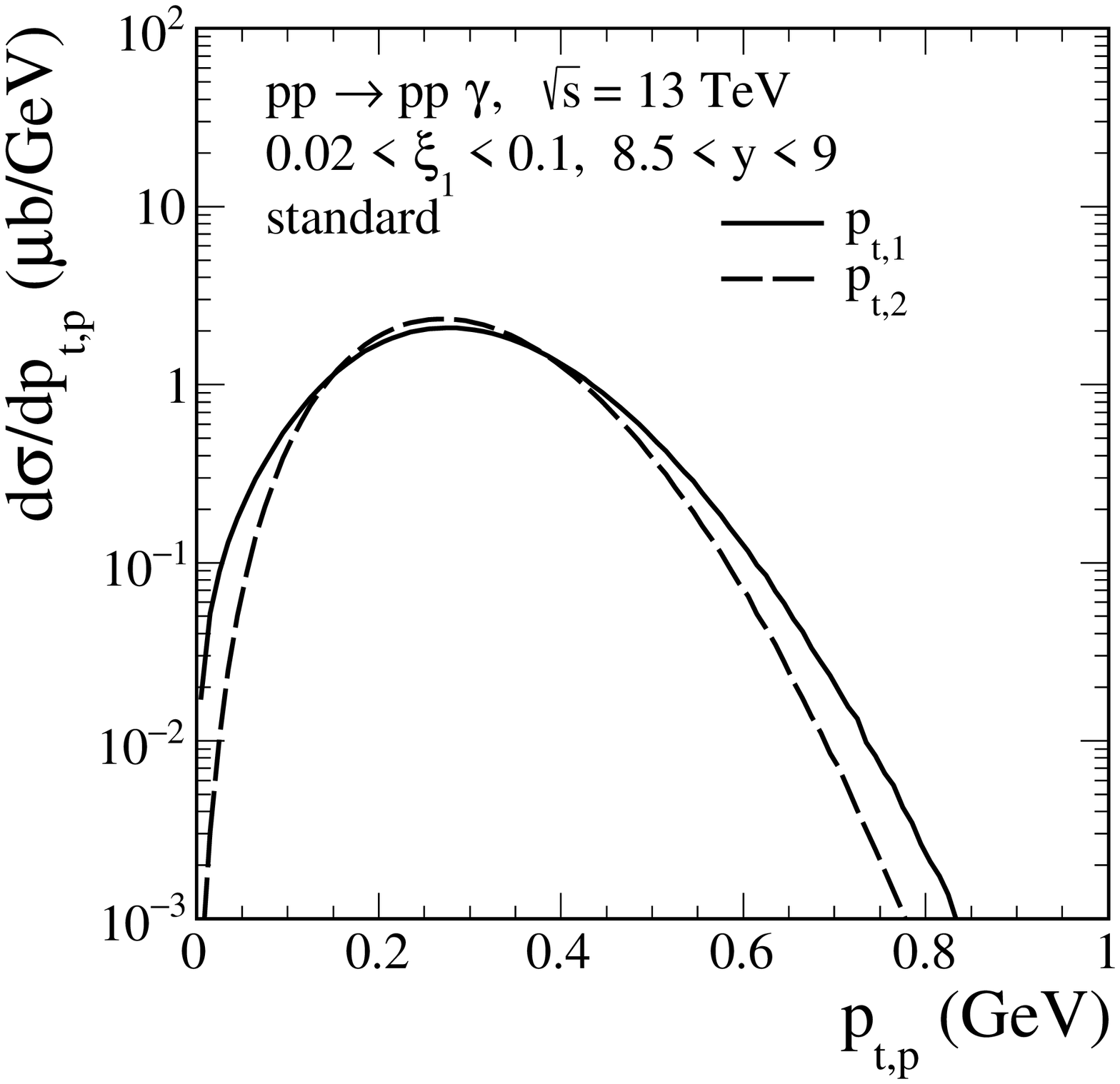}
\caption{\label{fig:2}
\small
The distributions in the absolute values of the transverse momenta 
of the outgoing protons for the $pp \to pp \gamma$ reaction
calculated for $\sqrt{s} = 13$~TeV,
$0.02 < \xi_{1} < 0.1$, $6 < {\rm y} < 13$ (left panel),
and $8.5 < {\rm y} < 9$ (right panel).
The solid (dashed) line, 
denoted as $p_{t,1}$ ($p_{t,2}$), 
corresponds to the proton $p(p_{1}')$ ($p(p_{2}')$).}
\end{figure}

In Fig.~\ref{fig:2} we show the distributions
in absolute value of the transverse momenta,
$p_{t,1}$ and $p_{t,2}$, of the outgoing protons
$p(p_{1}')$ and $p(p_{2}')$, 
see the solid and dashed lines, respectively.
Results for two ${\rm y}$ intervals are shown:
$6 < {\rm y} < 13$ (left panel) and $8.5 < {\rm y} < 9$ (right panel).
One can see that the cross sections
reach a maximum at $p_{t,p} \sim \sqrt{|t_{1,2}|} \sim 0.25$~GeV
and that $d\sigma/dp_{t,1} \neq d\sigma/dp_{t,2}$.
We find that for $p_{t,1} > 0.8$~GeV, 
the Pauli component gives a sizeable contribution.

Now we consider the azimuthal angles $\phi_{i}$
of the transverse momenta of the protons
$p(p_{i}')$
and the photon $\gamma(k)$
\begin{eqnarray}
&&\bhpi = p_{t,i} \,e^{i \phi_{i}}\,, \quad (i = 1,2)\,, \nonumber \\
&&\khpi = k_{\perp} \,e^{i \phi_{3}}\,,  \nonumber \\
&& 0 \leqslant \phi_{i} < 2 \pi\,,
\quad (i = 1,2,3)\,.
\label{phi_i}
\end{eqnarray}
Here the azimuth $\phi = 0$ corresponds
to some fixed transverse direction in the LHC system
which is also the c.m. system for our reactions.
Transverse-momentum conservation requires
\begin{eqnarray}
\bhpii + \bhpjj + \khpi = 0\,.
\label{3.1a}
\end{eqnarray}
Therefore, a measurement of $\bhpii$ and $\khpi$ determines also $\bhpjj$.

\begin{figure}[!ht]
\includegraphics[width=0.44\textwidth]{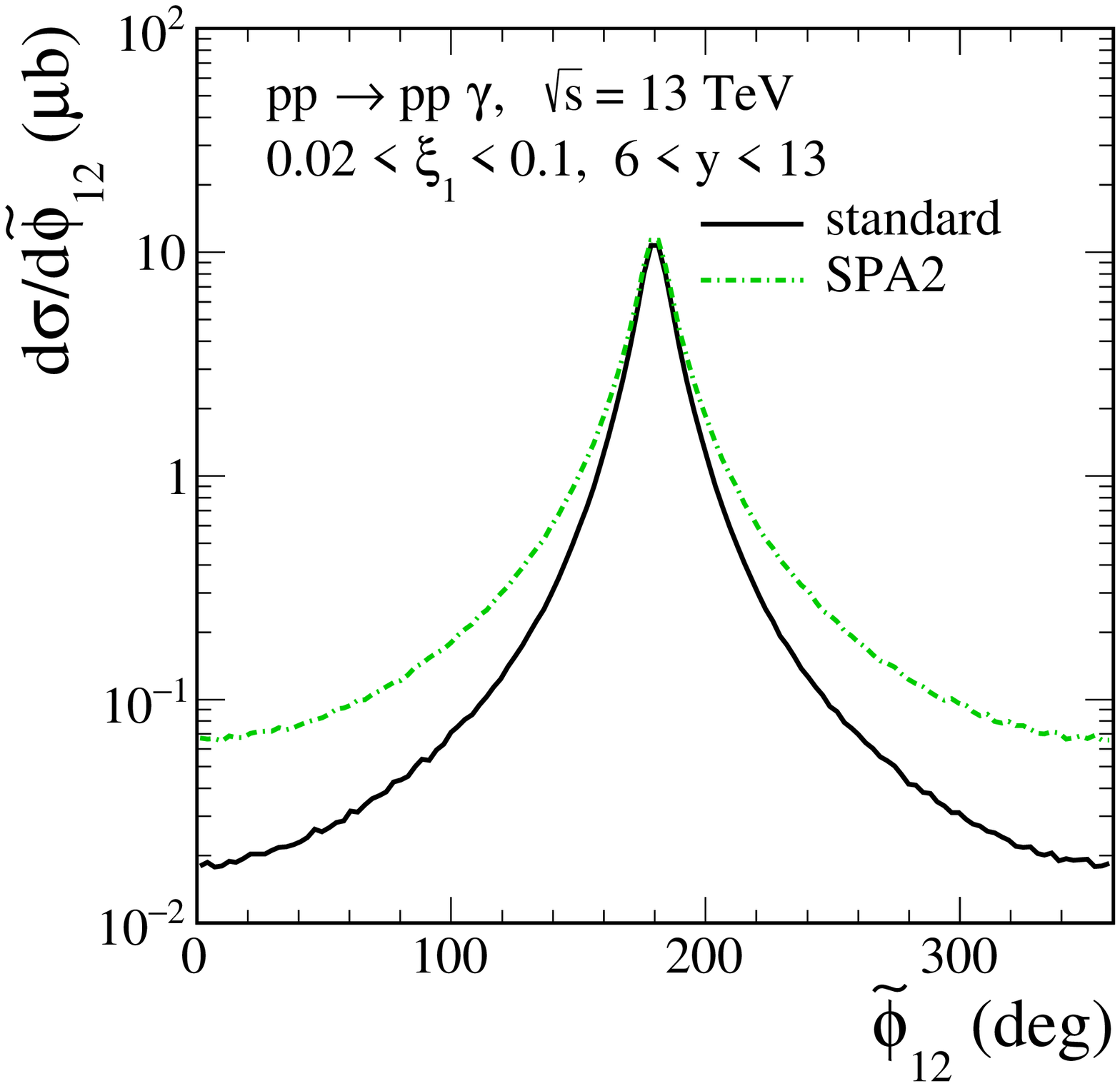}
\includegraphics[width=0.44\textwidth]{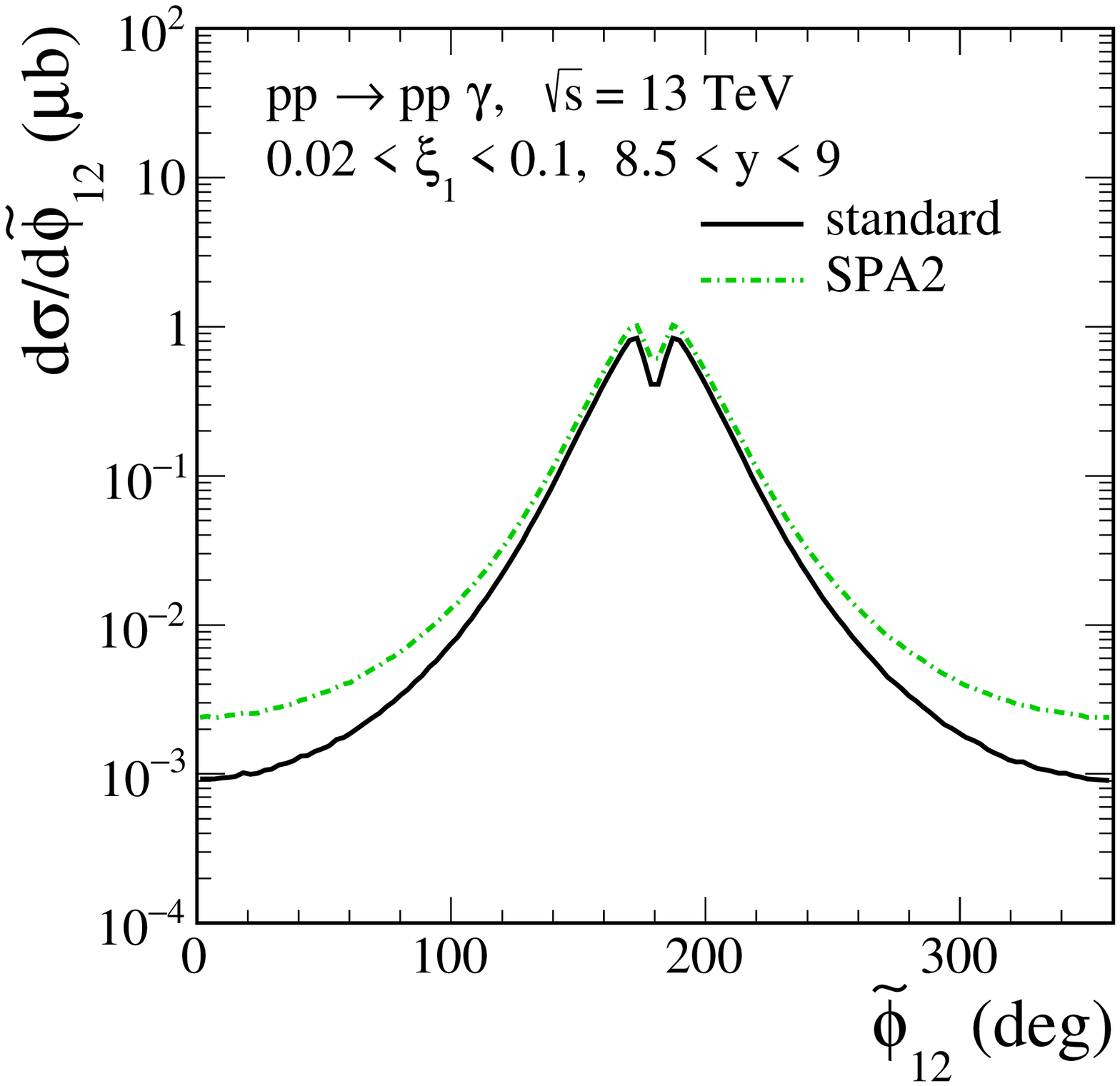}
\includegraphics[width=0.44\textwidth]{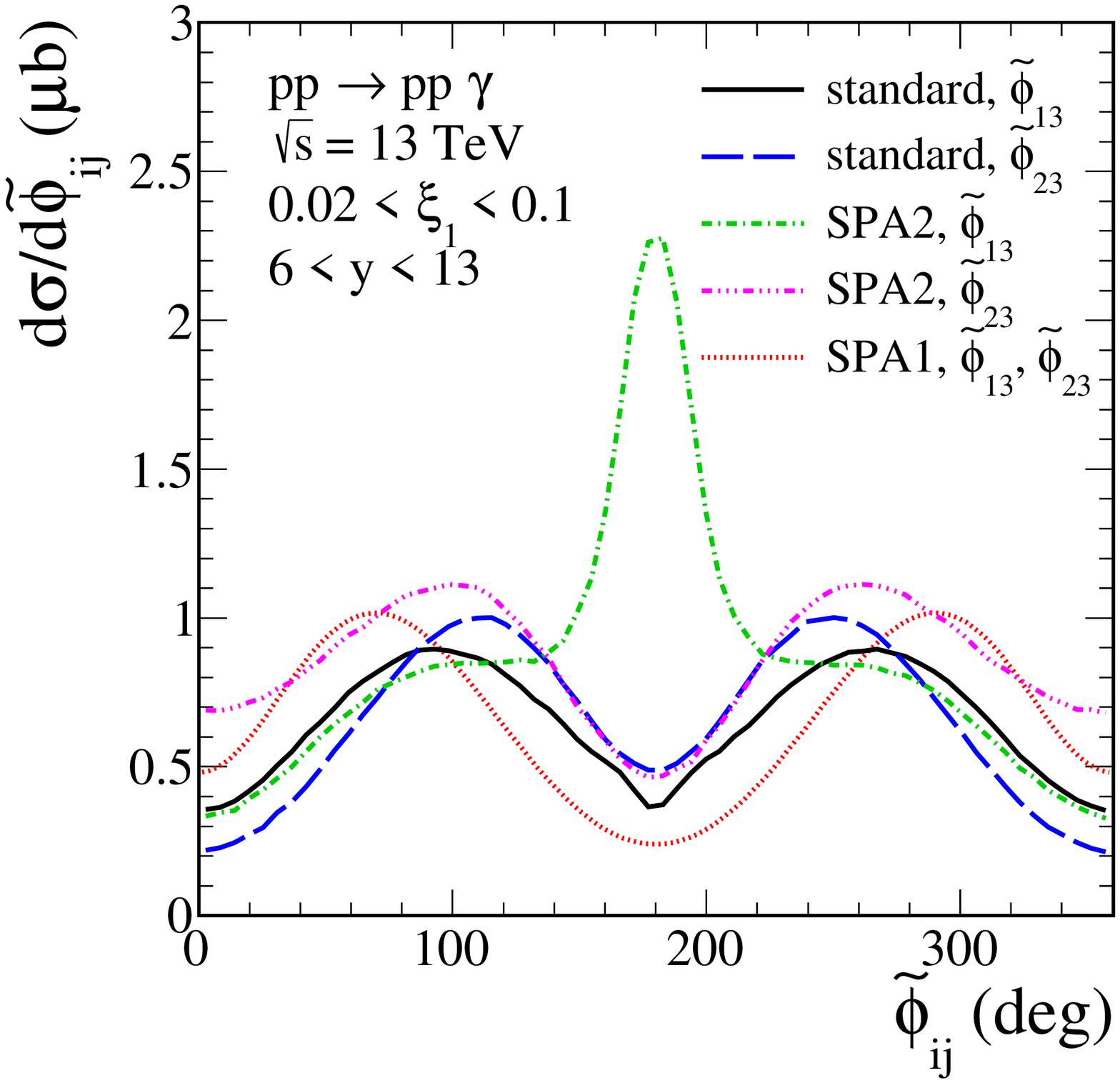}
\includegraphics[width=0.44\textwidth]{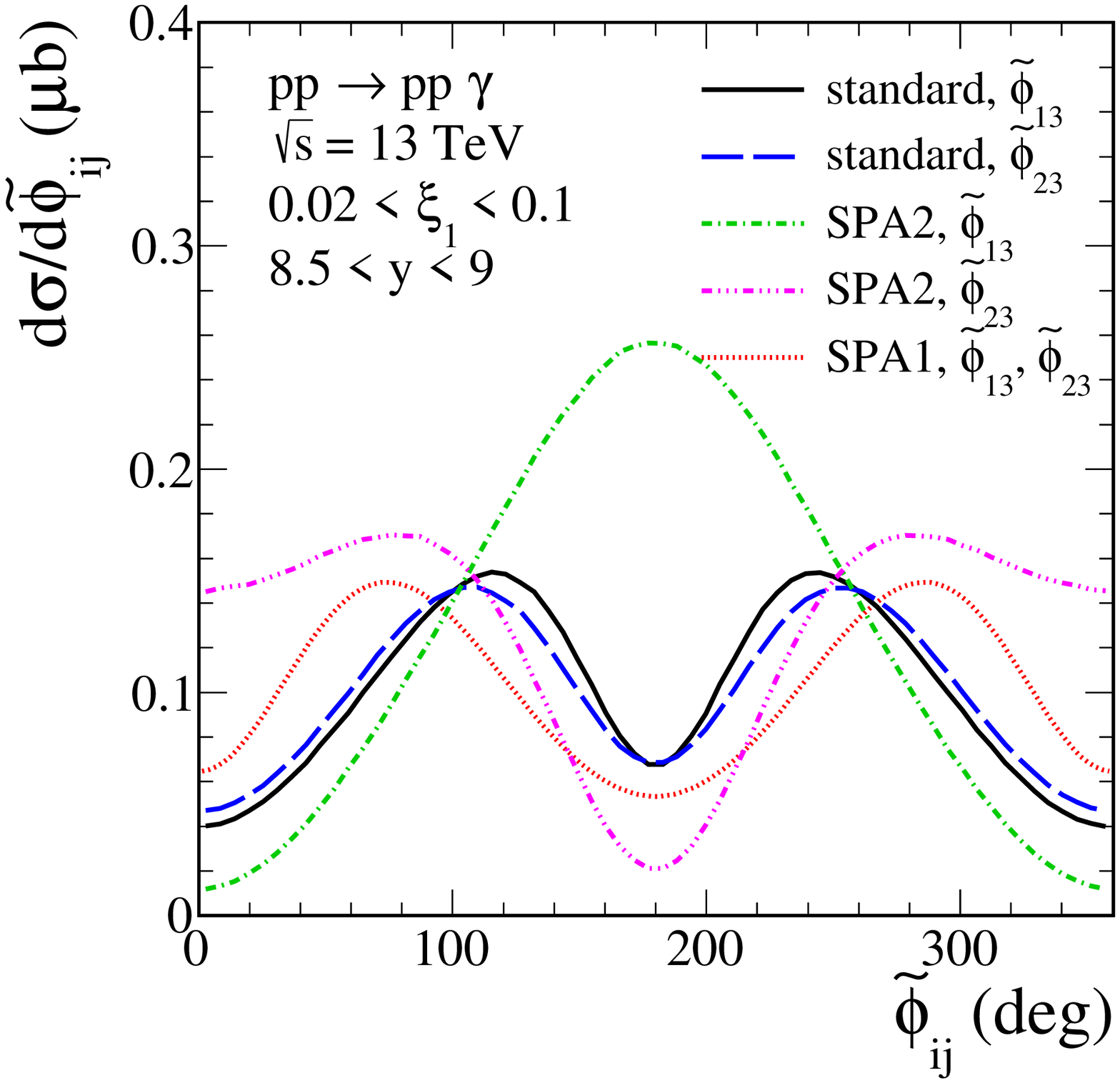}
\caption{\label{fig:3}
\small
The distributions in the angles 
$\tilde{\phi}_{ij}$ defined by (\ref{phi_ij})
for the $pp \to pp \gamma$ reaction
calculated for $\sqrt{s} = 13$~TeV,
$0.02 < \xi_{1} < 0.1$, $6 < {\rm y} < 13$ (left panel),
and $8.5 < {\rm y} < 9$ (right panel).
The results for the standard bremsstrahlung model and the SPAs 
are shown.}
\end{figure}

Figure~\ref{fig:3} shows the distributions in 
$\tilde{\phi}_{ij}$ defined as 
\begin{eqnarray}
\tilde{\phi}_{ij} = \phi_{i} - \phi_{j} \quad {\rm mod}(2 \pi)\,,
\label{phi_ij}
\end{eqnarray}
where we require
\begin{eqnarray}
0 \leqslant \tilde{\phi}_{ij} < 2 \pi\,.
\label{3.2a}
\end{eqnarray}
In the top panels of Fig.~\ref{fig:3} we show
the results in $\tilde{\phi}_{12}$,
the angle between 
the transverse momenta of the outgoing protons,
for our standard and SPA2 calculations.
For SPA1 (not shown here)
the outgoing protons are back-to-back 
($\tilde{\phi}_{12} = \pi$).
We see that also for our standard approach
$\tilde{\phi}_{12} \approx \pi$ gives the main contribution.
That is, very roughly the outgoing protons and the beam are
in one plane ${\cal S}_{0}$.
The width of the distribution in $\tilde{\phi}_{12}$
depends on the ${\rm y}$ cut with the more restrictive ${\rm y}$ cut
(right panel) giving a wider distribution.

The bottom panels of Fig.~\ref{fig:3} show the distributions
in $\tilde{\phi}_{13}$ ($\tilde{\phi}_{23}$),
the azimuthal angles between the proton $p(p_{1}')$ ($p(p_{2}')$)
and $\gamma(k)$; see Eq.~(\ref{phi_ij}).
The SPA1 and our standard results show maxima 
for $\tilde{\phi}_{13}$ and $\tilde{\phi}_{23}$
around $\pi/2$ and $3\pi/2$.
This corresponds to emission of the photon in a plane ${\cal S}_{1}$
which is orthogonal to the plane ${\cal S}_{0}$ defined above.
The SPA2 results
for the $\tilde{\phi}_{13}$ and $\tilde{\phi}_{23}$
distributions deviate very significantly from our standard results.

\newpage

\begin{figure}[!ht]
\includegraphics[width=0.44\textwidth]{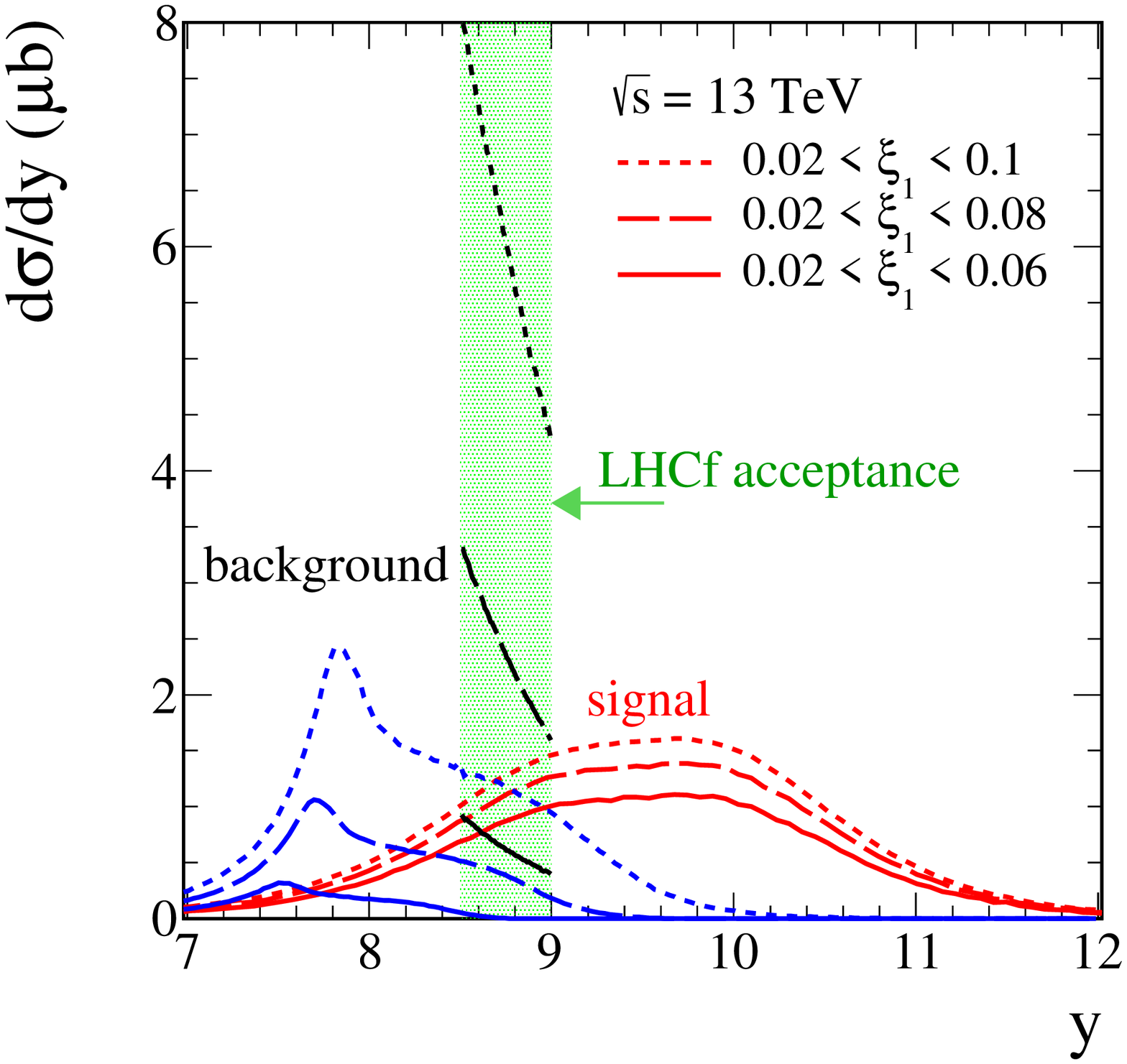}
\includegraphics[width=0.44\textwidth]{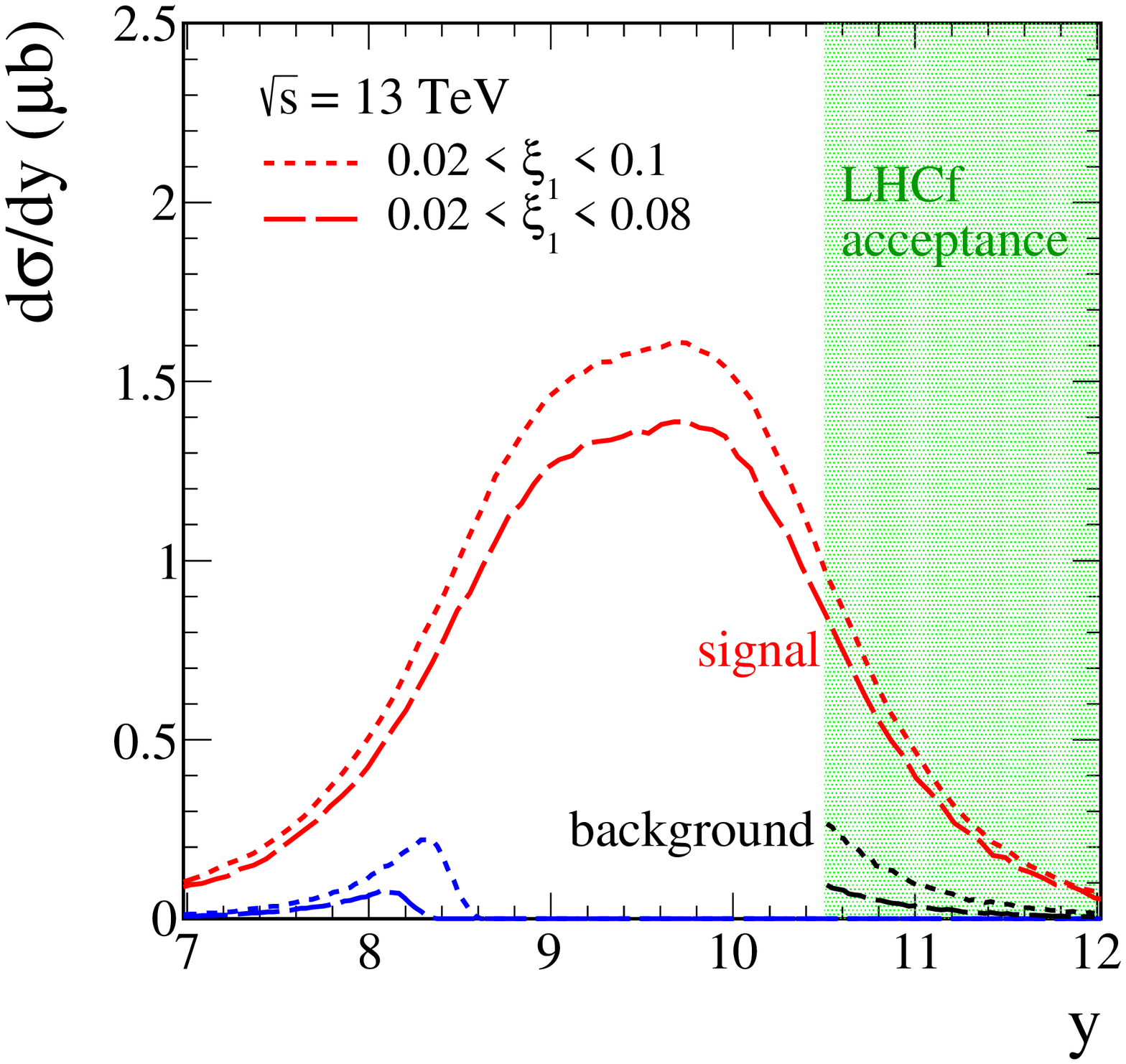}
\caption{\label{fig:background}
\small
The distribution in rapidity of the photon
for the signal (bremsstrahlung) and background contributions
for different ranges of $\xi_1$. 
The LHCf acceptance regions are marked by the green shaded areas.
In the left panel, the results within the LHCf acceptance
correspond to the acceptance region $8.5 < {\rm y} < 9$,
while for the right panel to ${\rm y} > 10.5$.
For the background contribution
we show the distributions of both photons from the decay of $\pi^0$.
The distributions of the first (measured) photon 
correspond to the black lines,
while the distributions of the second photon
correspond to the blue lines.
The dashed, long dashed, and full lines correspond to
the three $\xi_1$ intervals as indicated in the figure legends.}
\end{figure}

In Fig.~\ref{fig:background} we present the result of a study
of the $p p \to p p (\pi^0 \to \gamma \gamma)$ background.
For this we take the upper estimate
of the cross section for the $p p \to p p \pi^0$ reaction
(Drell-Hiida-Deck type model)
from \cite{Lebiedowicz:2013vya}
which corresponds to $\Lambda_{N} = \Lambda_{\pi} = 1$~GeV,
and without taking into account the absorptive corrections.\footnote{
One can see from Fig.~6 of \cite{Lebiedowicz:2013vya}
that at ${\rm y}_{\pi^{0}} \simeq 9$
the $\pi$-exchange mechanism
is the dominant background contribution.
The two other contributions, 
the proton exchange and the direct production of $\pi^{0}$,
are similar in magnitude but opposite in sign and 
thus cancel each other.
They can win over the $\pi$-exchange contribution for 
${\rm y}_{\pi^{0}} \gtrsim 10$.}
The red lines represent the distributions of 
our (signal) standard bremsstrahlung of a single photon 
associated with a proton for which the fractional energy loss $\xi_1$ 
is in the intervals specified in the figure legend.
We vary the upper limit of $\xi_1$. 
The signal distribution moderately depends on the upper limit;
see Fig.~\ref{fig:1}.
In the left panel of Fig.~\ref{fig:background},
for the background contribution we assume that
one photon is measured 
(called ``first'' in the following) by the LHCf
in the rapidity interval $8.5 < {\rm y} < 9$, 
$\omega > 130$~GeV, and with
$\omega_{\rm max} \approx \frac{\sqrt{s}}{2}\xi_{1, \rm max}$.
The distributions of the measured photon 
correspond to the black lines (within the LHCf acceptance),
while the distributions of the second photon
correspond to the blue lines.
In the latter case,
the maximum of the cross section corresponding to the background
is located outside the LHCf acceptance region.
The percentage of measured/unmeasured photons depends on 
the upper limit of $\xi_1$. 
For $\xi_{1,\rm max}$ = 0.06 (the solid lines) 
the second photon practically cannot be measured.
The signal-to-background ratio for
$\xi_{1,\rm max}$ = 0.06 is somewhat larger than 1.
In the right panel of Fig.~\ref{fig:background},
we present the contribution of the background 
for the ${\rm y} > 10.5$ acceptance range.
In this case the second photon cannot be measured.
Here, the signal-to-background ratio is of order of 4
for $\xi_{1,\rm max}$ = 0.1,
and about 10 for $\xi_{1,\rm max}$ = 0.08.

\subsection{Results for emission of two photons}
\label{sec:double}

\begin{figure}[!ht]
\includegraphics[width=0.44\textwidth]{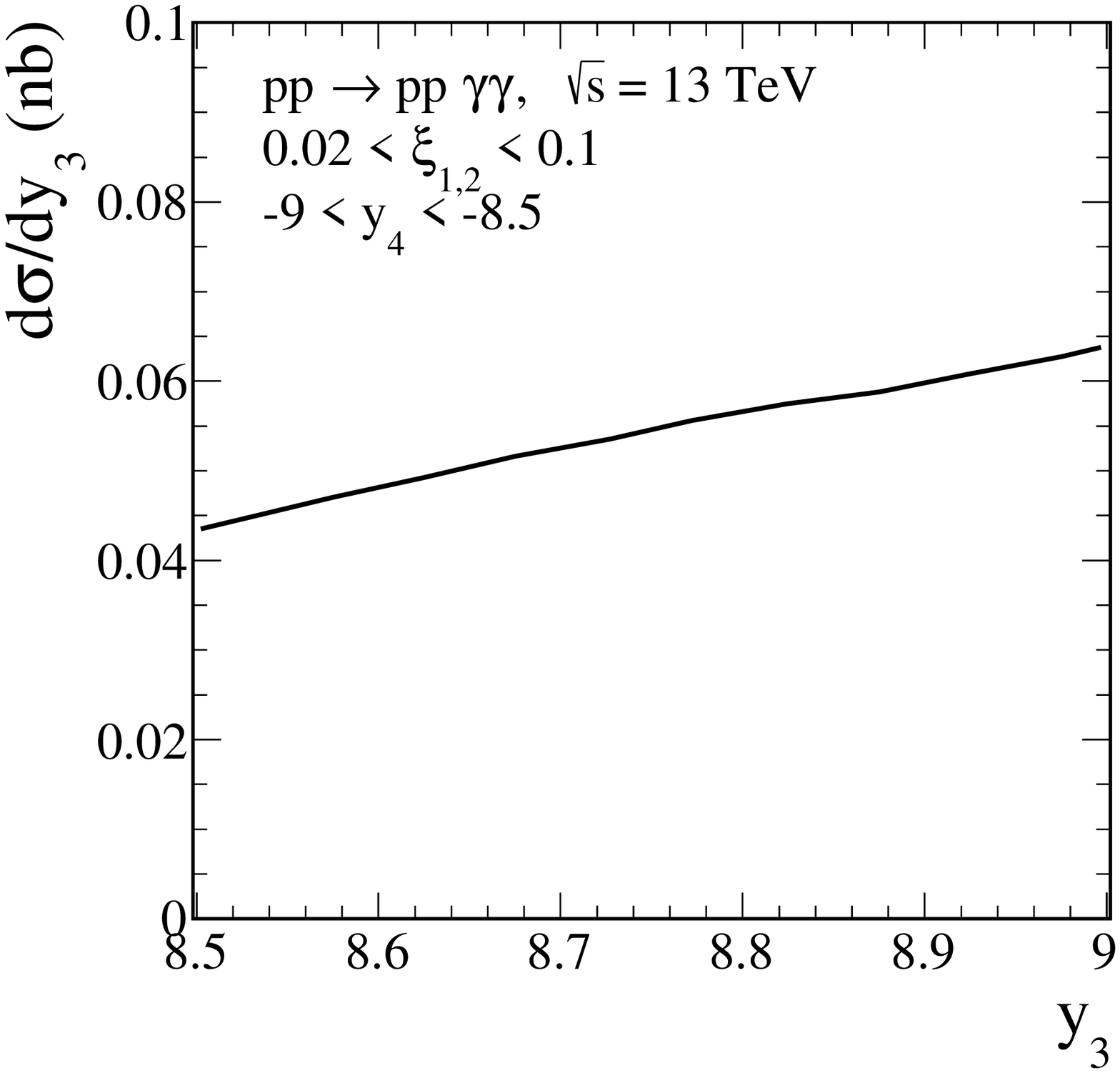}
\includegraphics[width=0.44\textwidth]{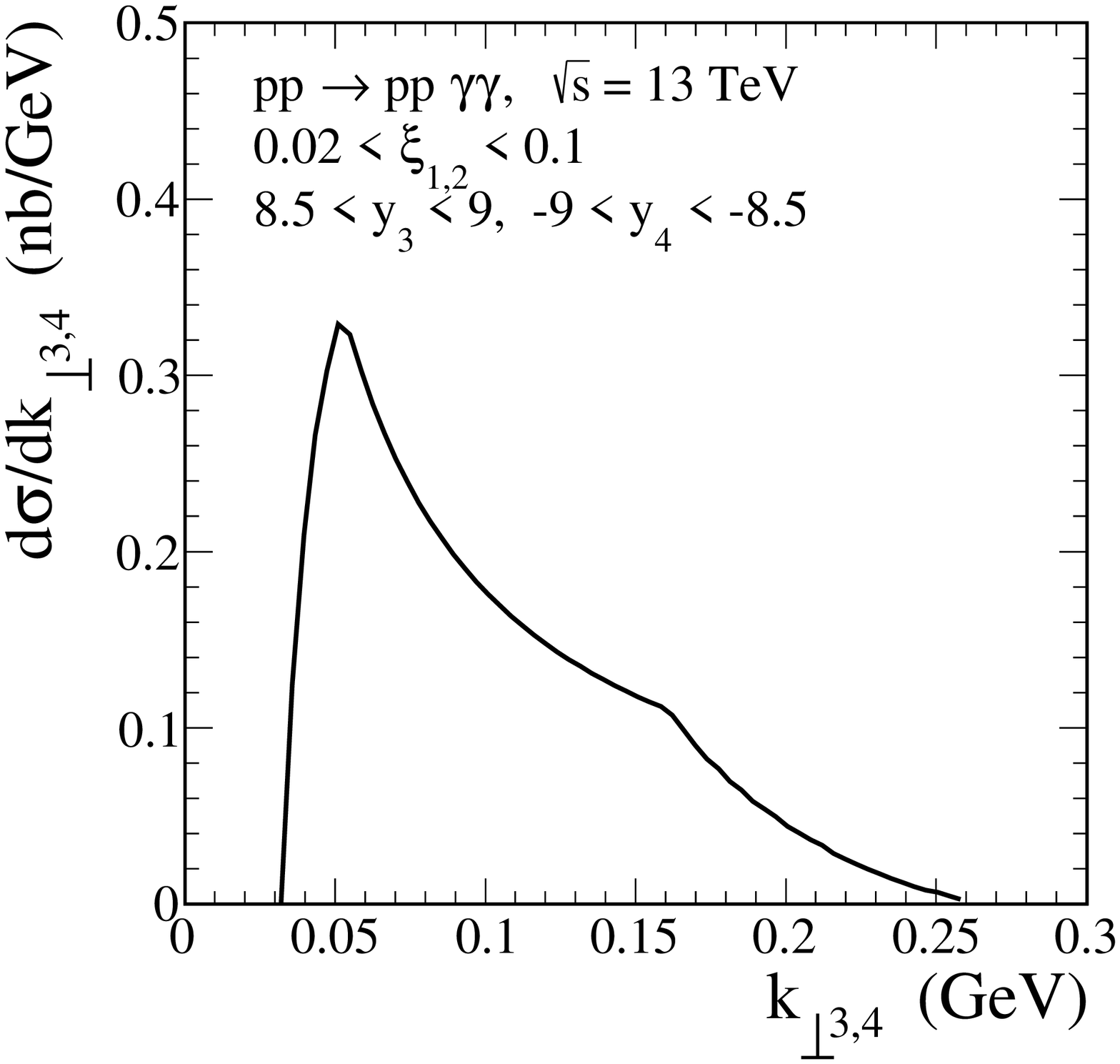}\\
\includegraphics[width=0.44\textwidth]{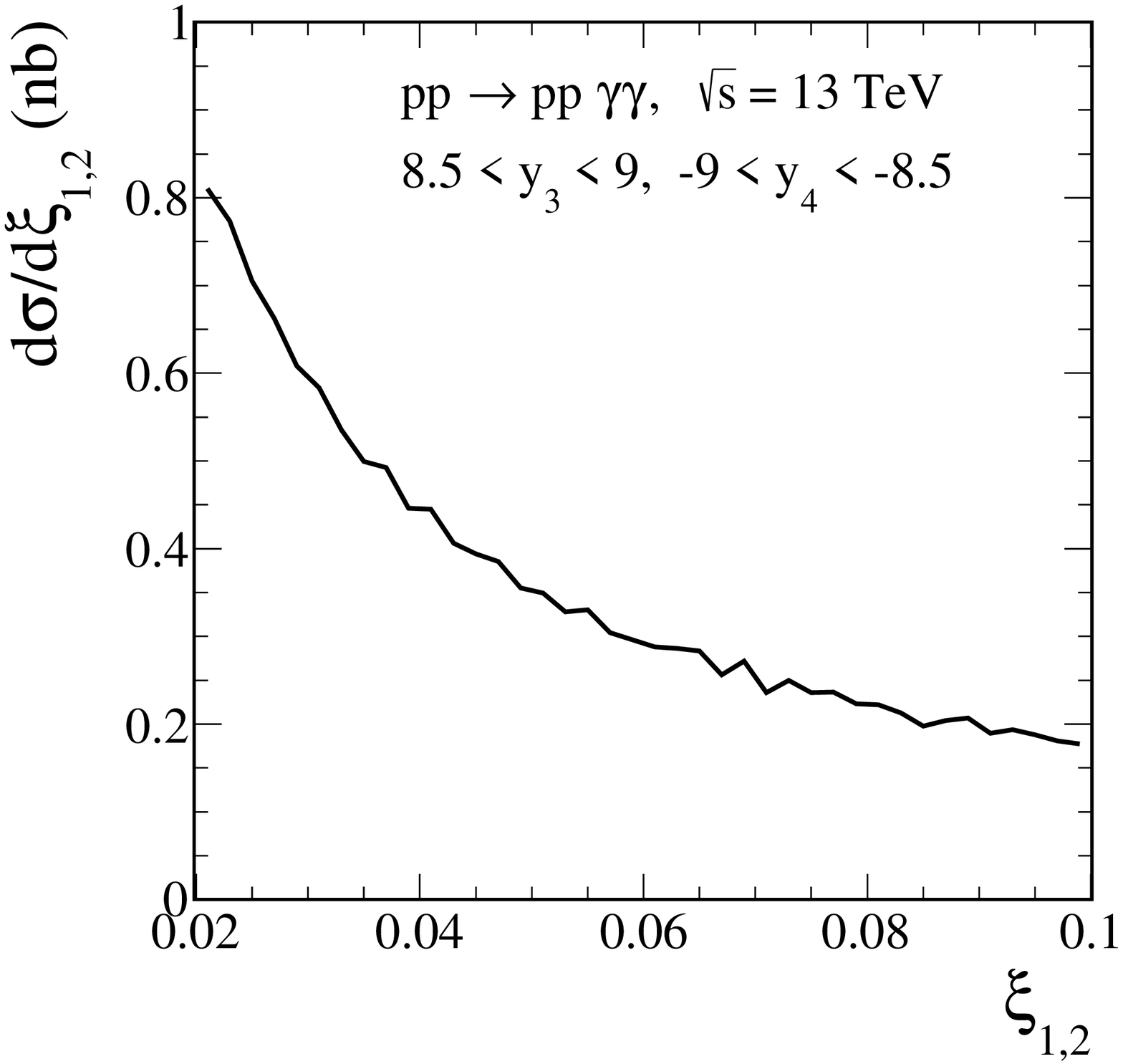}
\includegraphics[width=0.44\textwidth]{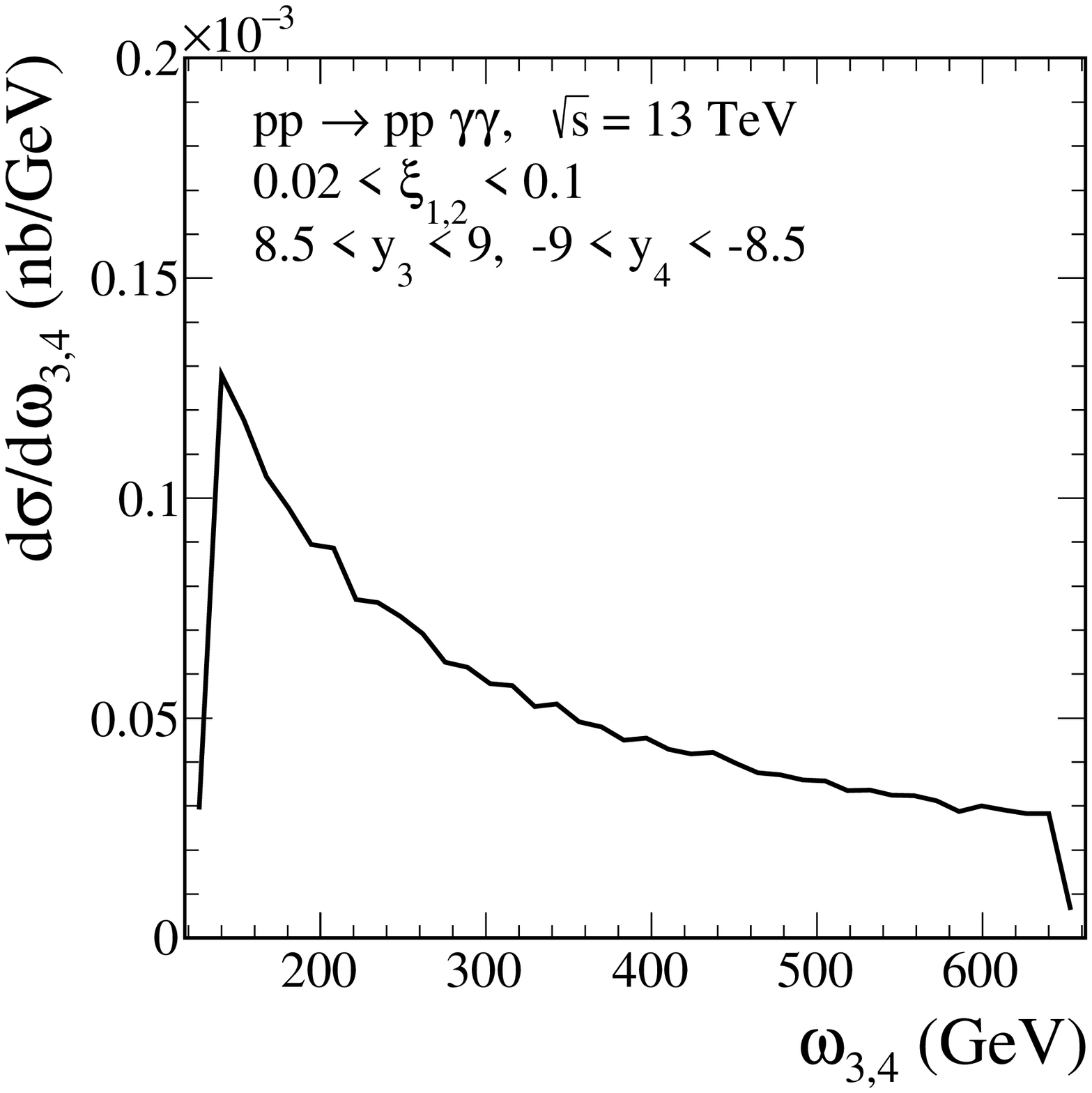}\\
\includegraphics[width=0.44\textwidth]{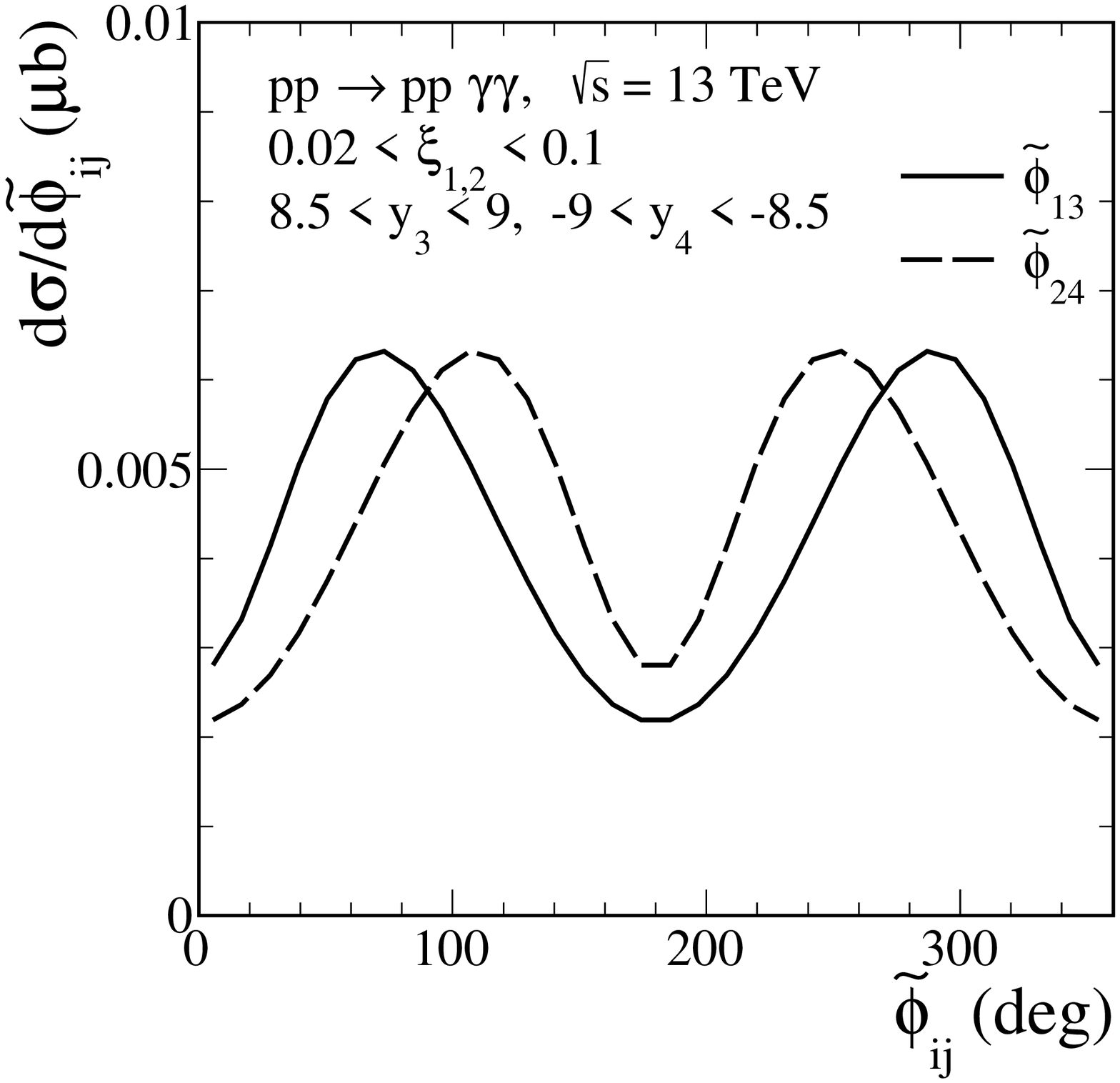}
\caption{\label{fig:10}
\small
The distributions for the two-photon bremsstrahlung
in the $pp \to pp \gamma \gamma$ reaction.
The calculations were done for $\sqrt{s} = 13$~TeV,
$8.5 < {\rm y}_{3} < 9$, $-9 < {\rm y}_{4} < -8.5$,
and $0.02 < \xi_{1,2} < 0.1$.
Shown are results for SPA1.}
\end{figure}

In Fig.~\ref{fig:10} we show the results 
for the $pp \to pp \gamma \gamma$ reaction,
discussed in Sec.~\ref{sec:2b},
calculated for $\sqrt{s} = 13$~TeV and in the kinematic region
specified in the figure caption.
The shapes of distributions in the first four panels are analogous
to the ones for single photon emission.
The lowest panel shows azimuthal correlations between forward proton
and forward photon (solid line) and between backward proton
and backward photon (dashed line).
The fact of seemingly different distributions for forward
and backward emissions is due to the way how the azimuthal angles
are defined in (\ref{phi_i}) and (\ref{phi_ij})
which leads to the symmetry relation
\begin{equation}
\frac{d \sigma^{\rm backward}}{d \phi}(\phi) = \begin{cases}
\frac{d \sigma^{\rm forward}}{d \phi}(\phi + \pi)& \mbox{for}\;\; 0 \leqslant \phi < \pi \,,\\
\frac{d \sigma^{\rm forward}}{d \phi}(\phi - \pi)& \mbox{for}\;\; \pi \leqslant \phi < 2 \pi \,.
 \end{cases}
\label{forward_backward_symmetry}
\end{equation}
This explains the observed differences between the solid
and dashed lines.

Finally we note that emission of the two photons
either from the $p_{a}$-$p_{1}'$ line 
or from the $p_{b}$-$p_{2}'$ line in Fig.~\ref{fig:pp_brems_gamgam_SPA1}
should essentially not contribute to the distributions
shown in Fig.~\ref{fig:10}.
Such processes will give two photons on one side
of the interaction point and
the final proton on the opposite side
will miss the $\xi$ cut which we impose.

\section{Conclusions}
\label{sec:4}

In this paper we have studied single- and double-photon bremsstrahlung
at very-forward and backward photon rapidities
in proton-proton collisions at high c.m. energies.
To calculate the amplitudes of the reactions
$pp \to pp \gamma$ and $pp \to pp \gamma\gamma$
the framework of the tensor-pomeron model was used.

We have started our analysis with the reaction $pp \to pp \gamma$,
single photon production.
We have compared our standard bremsstrahlung results
and the results using the approximations SPA1 and SPA2.
These approaches were discussed in our previous paper
\cite{Lebiedowicz:2022nnn} but in a different kinematic region.
In the present paper, we have shown that in the forward-rapidity region
and for $0.02 < \xi_{1} < 0.1$
(that corresponds to $130~{\rm GeV} < \omega < 650~{\rm GeV}$)
the standard results and the SPA1 results 
for purely photonic distributions 
($d\sigma/dk_{\perp}$, 
$d\sigma/d\omega$,
$d\sigma/d{\rm y}$)
are very close to each other, 
while the SPA2 overestimates differential cross sections; 
see Fig.~\ref{fig:1} and Fig.~\ref{fig:1_8.5_9}.

We have studied the azimuthal angle correlations
between outgoing particles (see Fig.~\ref{fig:3}).
We observe very interesting correlations between protons and photons.
Moreover, these correlations are significantly different
for our standard approach and for the SPA1 and SPA2 approaches.
Therefore, we emphasize that for detailed comparisons of our predictions
with experiment 
and in order to distinguish our standard and the approximate approaches
measurement of the outgoing protons would be
most welcome, if not indispensable.

We emphasize that for $\omega \to 0$ our bremsstrahlungs
distributions are an exact result of QCD plus lowest order electromagnetism.
This follows from Low's theorem \cite{Low:1958sn}
and the fact that our tensor-pomeron model describes proton-proton elastic
scattering at $\sqrt{s} = 13$~TeV quite well; see \cite{Lebiedowicz:2022nnn}.
But is $\omega$ in the range 130--650~GeV small enough?
In order to answer this question we have to consider,
for forward emission of the photon,
the scalar products $p_{a} \cdot k$ and $p_{1}' \cdot k$.
We get
\begin{eqnarray}
p_{a} \cdot k \approx p_{1}' \cdot k &\approx& 
\frac{\omega}{2 p_{a}^{0}} m_{p}^{2} + \frac{p_{a}^{0}}{2 \omega} k_{\perp}^{2} \nonumber \\
&\approx& \frac{1}{2} \xi_{1} m_{p}^{2} + \frac{1}{2} \frac{k_{\perp}^{2}}{\xi_{1}}
\,.
\label{4.1}
\end{eqnarray}
As we see from Fig.~\ref{fig:1}(b) the main part of the $k_{\perp}$ distribution
is for $k_{\perp} < 0.2$~GeV. 
With $0.02 < \xi_{1} < 0.1$ we have then always 
$p_{a} \cdot k \approx p_{1}' \cdot k < 1$~GeV$^{2}$.
Thus, we think that this is well in the region
where Low's theorem should be applicable.

We have estimated the coincidence cross section
for two-photon bremsstrahlung in the $pp \to pp \gamma \gamma$
reaction within the SPA1 approach.
We have required that the final state protons and photons
can be measured by the ATLAS forward proton spectrometers (AFP)
and LHCf detectors, respectively.
We have imposed the kinematical cuts
$8.5 < {\rm y}_{3} < 9$, $-9 < {\rm y}_{4} < -8.5$,
$0.02 < \xi_{1,2} < 0.1$,
and obtained the corresponding cross section 
$\sigma \simeq 0.03$~nb for $\sqrt{s} = 13$~TeV.
Our predictions can be verified by the ATLAS-LHCf measurement.

We have also briefly estimated the background contribution
due to the $p p \to p p \pi^0$ diffractive process for single
photon bremsstrahlung. We have compared the signal and background
contributions in two LHCf acceptance regions,
$8.5 < {\rm y} < 9$ and ${\rm y} > 10.5$.
One can increase the signal-to-background ratio to about 1
for the first acceptance region. For the second acceptance region the
ratio is bigger than 3.5.
We conclude that there is a chance to measure single photon
bremsstrahlung with the present experimental configuration discussed here.
For the two-photon bremsstrahlung a background estimate
is much more complicated
and goes beyond the scope of this Letter.

Let us finally comment on experimental signatures for our processes.
The single photon bremsstrahlung mechanism should be identifiable
by the measurement of proton and photon on one side and by
checking the exclusivity condition (no particles in the main detector)
without explicit measurement of the opposite side proton by AFP.
Whether this is sufficient requires further studies,
since such a measurement will probably include one-side diffractive
dissociation, which can be of the order of 20-30 \%.
Let us note that in order to isolate our signal reaction
$pp \to pp \gamma$ it would be very helpful if the transverse momenta
of the outgoing photon and protons could be measured.
For the signal reaction these transverse momenta must add up to zero;
see (\ref{3.1a}).
The background reactions discussed in Sec.~\ref{sec:single} and
above will not satisfy (\ref{3.1a}).
Thus, a cut on the quantity (\ref{3.1a}) could be used to eliminate
background to a good part.
The cross section for two photons on different sides is rather small
but should be measurable. 
Here a study of the background contributions should be done.

\section*{Acknowledgments}
We are very much indebted to Oldrich Kepka
for useful discussions on the possible ATLAS-LHCf measurement,
using AFP and LHCf detectors on one and both sides of the ATLAS interaction point.
This work was partially supported by
the Polish National Science Centre under Grant
No. 2018/31/B/ST2/03537
and by the Center for Innovation and Transfer of Natural Sciences and
Engineering Knowledge in Rzesz{\'o}w. 


\begin{thebibliography}{99}

\bibitem{Scholten:2002jb}
O.~Scholten and A.~Y. Korchin, {\em {Virtual-pion and two-photon production in
  $pp$ scattering},} \href{http://dx.doi.org/10.1103/PhysRevC.65.054004}{Phys.
  Rev. C {\bfseries 65} (2002) 054004},
  \href{http://arxiv.org/abs/nucl-th/0203078}{{arXiv:nucl-th/0203078}}.

\bibitem{Haberzettl:2010cg}
H.~Haberzettl and K.~Nakayama, {\em {Gauge-invariant formulation of $N N \to N
  N \gamma$},} \href{http://dx.doi.org/10.1103/PhysRevC.85.064001}{Phys. Rev. C
  {\bfseries 85} (2012) 064001},
\href{http://arxiv.org/abs/1011.1927}{{arXiv:1011.1927 [nucl-th]}}.

\bibitem{Korchin:1995ys}
A.~Y. Korchin and O.~Scholten, {\em {Dilepton production in nucleon-nucleon
  collisions and the low-energy theorem},}
  \href{http://dx.doi.org/10.1016/0375-9474(94)00457-X}{Nucl. Phys. A
  {\bfseries 581} (1995) 493}.

\bibitem{Korchin:1996up}
A.~Y. Korchin, O.~Scholten, and D.~van Neck, {\em {Low-energy theorems for
  virtual nucleon-nucleon bremsstrahlung; formalism and results},}
  \href{http://dx.doi.org/10.1016/0375-9474(96)00095-4}{Nucl. Phys. A
  {\bfseries 602} (1996) 423}.

\bibitem{Chwastowski:2015mua}
J.~Chwastowski, L.~Fulek, R.~Kycia, R.~Sikora, J.~Turnau, A.~Cyz, and
  B.~Pawlik, {\em {Feasibility Studies of Exclusive Diffractive Bremsstrahlung
  Measurement at RHIC Energies},}
  \href{http://dx.doi.org/10.5506/APhysPolB.46.1979}{Acta Phys. Polon. B
  {\bfseries 46} no.~10, (2015) 1979},
  \href{http://arxiv.org/abs/1501.06264}{{arXiv:1501.06264 [hep-ex]}}.

\bibitem{Chwastowski:2016jkl}
J.~J. Chwastowski, S.~Czekierda, R.~Kycia, R.~Staszewski, J.~Turnau, and
  M.~Trzebi{\'n}ski, {\em {Feasibility studies of the diffractive
  bremsstrahlung measurement at the LHC},}
  \href{http://dx.doi.org/10.1140/epjc/s10052-016-4181-y}{Eur. Phys. J. C
  {\bfseries 76} no.~6, (2016) 354},
  \href{http://arxiv.org/abs/1603.06449}{{arXiv:1603.06449 [hep-ex]}}.

\bibitem{Chwastowski:2016zzl}
J.~J. Chwastowski, S.~Czekierda, R.~Staszewski, and M.~Trzebi{\'n}ski, {\em
  {Diffractive bremsstrahlung at high-$\beta^\star$ LHC},}
  \href{http://dx.doi.org/10.1140/epjc/s10052-017-4789-6}{Eur. Phys. J. C
  {\bfseries 77} no.~4, (2017) 216},
  \href{http://arxiv.org/abs/1612.06066}{{arXiv:1612.06066 [hep-ex]}}.

\bibitem{Adriani:2022csq}
O.~Adriani {\em et~al.}, {\em {Measurement of forward photon production
  cross-section in $pp$ collisions at $\sqrt{s}$ = 510 GeV with RHICf
  detector},} \href{http://arxiv.org/abs/2203.15416}{{arXiv:2203.15416
  [hep-ex]}}.

\bibitem{LHCf:2012stt}
O.~Adriani {\em et~al.}, (LHCf Collaboration), {\em {Measurement of zero degree
  inclusive photon energy spectra for $\sqrt{s}= 900$ GeV proton-proton
  collisions at LHC},}
  \href{http://dx.doi.org/10.1016/j.physletb.2012.07.065}{Phys. Lett. B
  {\bfseries 715} (2012) 298},
  \href{http://arxiv.org/abs/1207.7183}{{arXiv:1207.7183 [hep-ex]}}.

\bibitem{LHCf:2011hln}
O.~Adriani {\em et~al.}, (LHCf Collaboration), {\em {Measurement of zero degree
  single photon energy spectra for $\sqrt{s} = 7$ TeV proton-proton collisions
  at LHC},} \href{http://dx.doi.org/10.1016/j.physletb.2011.07.077}{Phys. Lett.
  B {\bfseries 703} (2011) 128},
  \href{http://arxiv.org/abs/1104.5294}{{arXiv:1104.5294 [hep-ex]}}.

\bibitem{LHCf:2017fnw}
O.~Adriani {\em et~al.}, (LHCf Collaboration), {\em {Measurement of forward
  photon production cross-section in proton-proton collisions at $\sqrt{s}$ =
  13 TeV with the LHCf detector},}
  \href{http://dx.doi.org/10.1016/j.physletb.2017.12.050}{Phys. Lett. B
  {\bfseries 780} (2018) 233},
  \href{http://arxiv.org/abs/1703.07678}{{arXiv:1703.07678 [hep-ex]}}.

\bibitem{ATLAS:2017rme}
(ATLAS Collaboration), {\em {Measurement of contributions of diffractive
  processes to forward photon spectra in $pp$ collisions at $\sqrt{s} = 13$
  TeV},} ATLAS-CONF-2017-075.

\bibitem{Tiberio:2022aej}
A.~Tiberio {\em et~al.}, {\em {LHCf Run II physics results in proton-proton
  collisions at $\sqrt{s}$ = 13 TeV},}
  \href{http://dx.doi.org/10.22323/1.414.0121}{PoS {\bfseries ICHEP2022} (2022)
  121}.

\bibitem{LHCf:2022nbp}
H.~Menjo {\em et~al.}, (on behalf of the LHCf and RHICf Collaboration), {\em
  {Status and Prospects of the LHCf and RHICf experiments},}
  \href{http://dx.doi.org/10.22323/1.395.0301}{PoS {\bfseries ICRC2021} (2022)
  301}.

\bibitem{Ewerz:2013kda}
C.~Ewerz, M.~Maniatis, and O.~Nachtmann, {\em {A Model for Soft High-Energy
  Scattering: Tensor Pomeron and Vector Odderon},}
  \href{http://dx.doi.org/http://dx.doi.org/10.1016/j.aop.2013.12.001}{Annals
  Phys. {\bfseries 342} (2014) 31},
\href{http://arxiv.org/abs/1309.3478}{{arXiv:1309.3478 [hep-ph]}}.

\bibitem{Lebiedowicz:2021byo}
P.~Lebiedowicz, O.~Nachtmann, and A.~Szczurek, {\em {High-energy $\pi \pi$
  scattering without and with photon radiation},}
  \href{http://dx.doi.org/10.1103/PhysRevD.105.014022}{Phys. Rev. D {\bfseries
  105} no.~1, (2022) 014022},
  \href{http://arxiv.org/abs/2107.10829}{{arXiv:2107.10829 [hep-ph]}}.

\bibitem{Lebiedowicz:2022nnn}
P.~Lebiedowicz, O.~Nachtmann, and A.~Szczurek, {\em {Soft-photon radiation in
  high-energy proton-proton collisions within the tensor-Pomeron approach:
  Bremsstrahlung},} \href{http://dx.doi.org/10.1103/PhysRevD.106.034023}{Phys.
  Rev. D {\bfseries 106} no.~3, (2022) 034023},
  \href{http://arxiv.org/abs/2206.03411}{{arXiv:2206.03411 [hep-ph]}}.

\bibitem{Lebiedowicz:2023mhe}
P.~Lebiedowicz, O.~Nachtmann, and A.~Szczurek, {\em {Central exclusive
  diffractive production of a single photon in high-energy proton-proton
  collisions within the tensor-Pomeron approach},}
  \href{http://dx.doi.org/10.1103/PhysRevD.107.074014}{Phys.
  Rev. D {\bfseries 107} (2023) 074014},
  \href{http://arxiv.org/abs/2302.07192}{{arXiv:2302.07192 [hep-ph]}}.

\bibitem{Khoze:2010jv}
V.~A. Khoze, J.~W. L{\"a}ms{\"a}, R.~Orava, and M.~G. Ryskin, {\em {Forward
  physics at the LHC: detecting elastic $pp$ scattering by radiative photons},}
  \href{http://dx.doi.org/10.1088/1748-0221/6/01/P01005}{JINST {\bfseries 6}
  (2011) P01005},
\href{http://arxiv.org/abs/1007.3721}{{arXiv:1007.3721 [hep-ph]}}.

\bibitem{Lebiedowicz:2013xlb}
P.~Lebiedowicz and A.~Szczurek, {\em {Exclusive diffractive photon
  bremsstrahlung at the LHC},}
  \href{http://dx.doi.org/10.1103/PhysRevD.87.114013}{Phys. Rev. D {\bfseries 87}
  (2013) 114013},
\href{http://arxiv.org/abs/1302.4346}{{arXiv:1302.4346 [hep-ph]}}.

\bibitem{Khoze:2017igg}
V.~A. Khoze, A.~D. Martin, and M.~G. Ryskin, {\em {Can invisible objects be
  'seen' via forward proton detectors at the LHC?},}
  \href{http://dx.doi.org/10.1088/1361-6471/aa6457}{J. Phys. G {\bfseries 44}
  no.~5, (2017) 055002},
  \href{http://arxiv.org/abs/1702.05023}{{arXiv:1702.05023 [hep-ph]}}.

\bibitem{Cisek:2011vt}
A.~Cisek, P.~Lebiedowicz, W.~Sch{\"a}fer, and A.~Szczurek, {\em {Exclusive
  production of $\omega$ meson in proton-proton collisions at high energies},}
  \href{http://dx.doi.org/10.1103/PhysRevD.83.114004}{Phys. Rev. D {\bfseries 83}
  (2011) 114004},
\href{http://arxiv.org/abs/1101.4874}{{arXiv:1101.4874 [hep-ph]}}.

\bibitem{Lebiedowicz:2013vya}
P.~Lebiedowicz and A.~Szczurek, {\em {Exclusive $p p \to p p \pi^{0}$ reaction
  at high energies},} \href{http://dx.doi.org/10.1103/PhysRevD.87.074037}{Phys.
  Rev. D {\bfseries 87} (2013) 074037},
\href{http://arxiv.org/abs/1303.2882}{{arXiv:1303.2882 [hep-ph]}}.

\bibitem{Foroughi-Abari:2021zbm}
S.~Foroughi-Abari and A.~Ritz, {\em {Dark sector production via proton
  bremsstrahlung},} \href{http://dx.doi.org/10.1103/PhysRevD.105.095045}{Phys.
  Rev. D {\bfseries 105} no.~9, (2022) 095045},
  \href{http://arxiv.org/abs/2108.05900}{{arXiv:2108.05900 [hep-ph]}}.

\bibitem{Feng:2022inv}
J.~L. Feng {\em et~al.}, {\em {The Forward Physics Facility at the
  High-Luminosity LHC},} \href{http://dx.doi.org/10.1088/1361-6471/ac865e}{J.
  Phys. G {\bfseries 50} no.~3, (2023) 030501},
  \href{http://arxiv.org/abs/2203.05090}{{arXiv:2203.05090 [hep-ex]}}.

\bibitem{Trzebinski:2014vha}
M.~Trzebi\'nski, {\em {Machine Optics Studies for the LHC Measurements},}
  \href{http://dx.doi.org/10.1117/12.2074647}{Proc. SPIE Int. Soc. Opt. Eng.
  {\bfseries 9290} (2014) 929026},
  \href{http://arxiv.org/abs/1408.1836}{{arXiv:1408.1836 [physics.acc-ph]}}.

\bibitem{Low:1958sn}
F.~E. Low, {\em {Bremsstrahlung of very low-energy quanta in elementary
  particle collisions},} \href{http://dx.doi.org/10.1103/PhysRev.110.974}{Phys.
  Rev. {\bfseries 110} (1958) 974}.
  
\end{thebibliography}

\end{document}